\documentclass{elsarticle}

\usepackage[utf8]{inputenc}
\usepackage[english]{babel}
\usepackage{graphicx}
\usepackage{amsmath, accents}
\usepackage{amssymb}
\usepackage{amsfonts}
\usepackage{amsthm}
\usepackage{mathrsfs}
\usepackage{latexsym}
\usepackage{subfigure}
\usepackage[ruled,vlined]{algorithm2e}
\usepackage{color}
\usepackage{xcolor}
\usepackage{enumitem}
\usepackage[pagewise]{lineno}

\newcommand{\Z}{\mathbb{Z}}
\newcommand{\R}{\mathbb{R}}
\newcommand{\abs}[1]{\left\vert #1\right\vert}

\theoremstyle{definition}
\newtheorem{teo}{Theorem}
\newtheorem{ex}[teo]{Example}
\newtheorem{defin}[teo]{Definition}
\newtheorem{prop}[teo]{Lemma}
\newtheorem{rem}[teo]{Remark}
\newtheorem{cor}[teo]{Corollary}

\journal{Discrete Applied Mathematics}

\begin{document}

\title{A rounding theorem for unique binary tomographic reconstruction}

\author[mi]{Paolo Dulio\corref{cor1}}
\ead{paolo.dulio@polimi.it}

\author[bs]{Silvia M.C.~Pagani}
\ead{silvia.pagani@unicatt.it}

\address[mi]{Dipartimento di Matematica ``F.~Brioschi'', Politecnico di Milano, Piazza Leonardo da Vinci $32$, I-$20133$ Milano, Italy}
\address[bs]{Dipartimento di Matematica e Fisica ``N.~Tartaglia'', Università Cattolica del Sacro Cuore, via Musei $41$, $25121$ Brescia, Italy}

\cortext[cor1]{Corresponding author}

\begin{abstract}
Discrete tomography deals with the reconstruction of images from projections collected along a few given directions. Different approaches can be considered, according to different models. In this paper we adopt the grid model, where pixels are lattice points with integer coordinates, X-rays are discrete lattice lines, and projections are obtained by counting the number of lattice points intercepted by X-rays taken in the assigned directions.

We move from a theoretical result that allows uniqueness of reconstruction in the grid with just four suitably selected X-ray directions. In this framework, the structure of the allowed ghosts is studied and described. This leads to a new result, stating that the unique binary solution can be explicitly and exactly retrieved from the minimum Euclidean norm solution by means of a rounding method based on some special entries, which are precisely determined. A corresponding iterative algorithm has been implemented, and tested on a few phantoms having different characteristics and structure.
\end{abstract}

\begin{keyword}
Binary tomography; discrete tomography; lattice direction; lattice grid; minimum norm solution; uniqueness of reconstruction.
\end{keyword}

\maketitle

\section{Introduction}
It is well known that a large class of tomographic problems concerns the reconstruction of an unknown object by means of partial data coming from its projections, collected by means of X-rays, and taken along given directions. Starting from the first scanner invented by Cormak and Hounsfield (1979 Nobel prize for Physiology or Medicine), who autonomously rediscovered and implemented the early theory of Radon (\cite{R}), technology has greatly improved throughout the years and has allowed tomography to be applied in several scientific areas, and exploiting different methodologies.

In the Radon approach angles under which projections are considered are available in the whole continuous interval $[0,\pi)$, and the radiation has good analytic properties. This allows the resulting filtered back-projection (FBP) inversion formula to be obtained by means of integration. However, in real applications, due to mechanical and physical reasons involved in the acquisition process, the typical assumptions of the continuous approach are not fulfilled, which can lead to a FBP reconstructed image of poor quality, due to the formation of artifacts and the presence of noise. This leads to look for different reconstruction algorithms, in particular of iterative nature (see for instance \cite{LOSSIBS}).

Since the scan devices allow to collect only a finite number of projections, along a finite set of directions having  rational slopes, the tomographic problem can be re-defined inside a lattice grid. In the typical frame of \emph{discrete tomography} (DT) \cite{HKbook99,HK1} only few types of different densities (say, 2-6) are involved. Density is assumed to be constant inside a same pixel of the resulting grid, so that the object to be reconstructed is shown under a finite resolution. In the special case of \emph{binary tomography} (BT) homogeneous objects are considered, and one is interested in detecting the presence or the absence of the object itself at different parts of the working grid.

The discrete modeling of the tomographic problem implies that there are no chances, in general, of achieving an exact reconstruction by the standard mathematical algorithms. Moreover, in some applications, the required number of directions, along which projections are taken, is very limited, in order to avoid damaging the objects to be studied. This leads to strong ambiguities in DT reconstructions (\cite{FLRS,FS}), and different approaches for a quantitative description of their uncertainty (see for instance \cite{GP,V1,V2,VNNB}) and stability (\cite{AB,AGT,V3}) have already been explored. As a consequence, we are mainly interested in looking for conditions that can limit the number of allowed solutions, and possibly for uniqueness conditions. Sometimes uniqueness results can be achieved by introducing some geometric conditions, such as convexity (\cite{GG}) or additivity (\cite{BDP3,BrPe,FS}).

When looking for efficient reconstruction algorithms, one should try to match some requirements. The first one is that the number of directions along which X-rays are performed cannot be too much large, in order to avoid huge amount of radiation. Due to dose constraints, this also reflects in the second request that, even for a small number of directions, the number of collected projections should be kept limited. The third desirable property is that the percentage of correctly reconstructed pixels should be high, so that, in principle, a reconstruction algorithm should be based on some a priori conditions that guarantee a limited number of tomographic reconstructions.

In this paper we focus, first of all, on the \emph{grid model}, largely employed in DT (see for instance \cite{GG,Ha-Ti}), where pixels are lattice points with integer coordinates, X-rays are discrete lattice lines, and projections are obtained by counting the number of lattice points intercepted by X-rays taken in the assigned directions. In \cite{VBN1}, projection dependency of the quality of tomographic outputs was investigated, by comparing reconstructions of a same phantom from different sets of directions. Analogously, we base on a theoretical uniqueness result for BT obtained in \cite{BDP1} (and generalized in \cite{BDP2} to higher dimensions), showing that exact noise-free binary reconstructions can be obtained in any grid with a suitable selection of just four directions, depending on the grid.

The tomographic problem can be modeled in terms of linear system of equations $A\mathbf{x}=\mathbf{p}$, where $A$ is the projection matrix, mapping an image $\mathbf{x}$ to a vector $\mathbf{p}$ of projection data, collected by means of X-rays in assigned directions. In general the linear system is highly under-determined, meaning that the reconstruction problem in the grid model is typically ill-posed (see for instance \cite{GGP, Ha-Ti, HKbook99, HK1}). Consequently, measures for testing the quality of a reconstruction w.r.t.~the unknown original image have been developed. In particular, the solution $\mathbf{x}^{\ast}$ having minimal Euclidean norm allows to bound the distance between different binary solutions (see \cite{BFHT,ht_siam,vdht}). This suggests that $\mathbf{x}^{\ast}$ should be considered as a kind of reference image in any binary tomographic reconstruction problem.

We recall that many combinatorial problems of interest can be encoded as integer linear programs, whose solution is in general NP-hard, and this is the case also for the tomographic reconstruction problem when the number of directions is greater than two. A usual strategy consists in relaxing the integer constraint into the real numbers. For instance, a binary problem where all variables are either $0$ or $1$ can be relaxed by requiring that each variable belongs to the real interval $[0,1]$. Then methods are employed in order to find the region of admissible solutions of the relaxed problem, where the optimal integer solutions should be sought. In particular, the optimal value could be obtained by integer rounding. This is the case for any integer programming problem satisfying the integer round-up property (IRUP), where the optimal value is provided by the nearest integer greater than, or equal to, the optimal value of the corresponding linear programming relaxation (\cite{BATR}). For instance, certain classes of cutting stock problems fulfill the IRUP (\cite{MAR}), even if it was shown (\cite{MAR1}) that the rounding property does not hold in general. This motivated the proposals of subsequent modifications of the IRUP (see for instance \cite{SCHTER,SCHTER1}) and further extensions to mixed integer linear programming, where only some of the involved variables are constrained to be integers. These problems are generally solved by using a branch-and-bound algorithm, based on the observation that the enumeration of integer solutions has a tree structure (see for instance the recent survey \cite{VIELMA}).

Moving from the above considerations, we follow the idea that integer solutions of the tomographic problem could be determined by integer rounding suitable solutions of the corresponding linear system $A\mathbf{x}=\mathbf{p}$. We relate the results in \cite{BFHT} with those in \cite{BDP1} to provide a method which allows to exactly reconstruct binary images from suitable sets of four directions. In particular we show (Theorem \ref{teo:rounding} and Corollary \ref{cor:roundingreconstruction}) that such sets guarantee the existence of a unique binary solution, which can be explicitly reconstructed from the minimum norm solution of the linear system.

The paper is organized as follows. In Section \ref{sec:preliminaries} we give the necessary preliminary definitions and notations. We recall the algebraic approach to DT and comment on the grid model. In Section \ref{sec:theoretical} we state the uniqueness theorem proved in \cite{BDP1} and a few further useful results for the construction of sets of directions ensuring binary uniqueness. In Section \ref{sec:ghosts} details are given concerning the structure of ghosts determined by sets of binary uniqueness. In Section \ref{sec:reconstruction}, the knowledge of the ghost sizes, combined with geometrical information concerning the real-valued solution of $A\mathbf{x}=\mathbf{p}$ having minimal Euclidean norm, leads to a binary rounding uniqueness theorem. The corresponding binary reconstruction algorithm (BRA) is presented in detail and its complexity is discussed. Moreover, BRA is applied on some phantoms taken from \cite{BFHT}. Section \ref{sec:conclusion} describes possible further work and concludes the paper.

\section{Preliminaries}\label{sec:preliminaries}
A \emph{(lattice) direction} is a pair $(a,b)$ of coprime integers such that $a=1$ if $b=0$ and conversely $b=1$ if $a=0$. We can assume, without loss of generality, that $a\geq0$. By \emph{lines with direction} $(a,b)\in\Z^{2}$ we mean lattice lines defined in the $x,y$ plane by equations of  the form $L:ay=bx+t$, where $t\in\Z$. Here we assume the horizontal axis to be oriented from left to right, and the vertical axis downwards. A finite subset of $\Z^2$ is said to be a {\em lattice set}. For a lattice set $E$, and a vector $u\in\Z^2$, we denote by $E+u$ the lattice set obtained by translating each point of $E$ along $u$. We are concerned with the reconstruction of \emph{binary images}, which can be represented as a finite set of points $E\subset\Z^2$, or as a function mapping each element, called \emph{pixel}, of the domain to either 0 or 1 (its \emph{value}). Sometimes we will say that we reconstruct a pixel instead of a binary image defined on that pixel.

\begin{figure}[htbp]
\centering
\subfigure[]
{\includegraphics[scale=1]{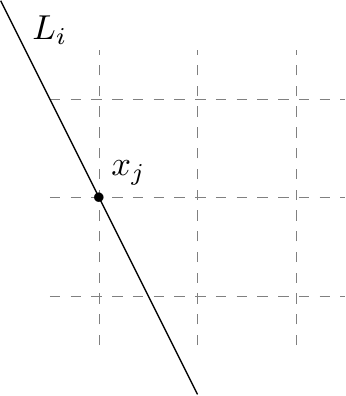} \label{fig:gridmodel}}
\hspace{1.5cm}
\subfigure[]
{\includegraphics[scale=1]{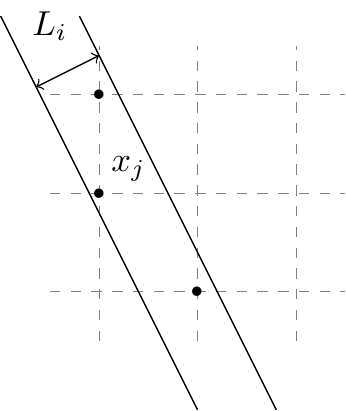} \label{fig:stripmodel}}
\caption{(a) The grid model. (b) The discrete strip model.}
\label{fig:models}
\end{figure}

We focus on the \emph{grid model}, where pixels are lattice points with integer coordinates, X-rays are discrete lattice lines \footnote{We remark that the term \textit{X-ray} is usually employed in DT as the measurement (see for instance \cite{GG}), not as the line intercepting the grid points. Here we prefer to adopt X-rays for lines, and to use the term \textit{projection} to denote the measure.} and projections are obtained by counting the number of lattice points intercepted by X-rays taken in the assigned directions (see Figure \ref{fig:gridmodel}). If we replace lattice lines with lattice strips we get a \emph{discrete strip model} (see Figure \ref{fig:stripmodel}). It is also known as \emph{Dirac model} (\cite{GEFE}), and differs from the discrete strip model considered in \cite{ZLYW}, where pixels are not collapsed in a lattice point, strips are continuous, and the contribution of a pixel to a given projection relates to the portion of pixel covered by the strip.

The discretization process of the continuous tomographic reconstruction method naturally  leads to an algebraic approach. Here, reconstructing an image from its projections is equivalent to solving the linear system
\begin{equation}\label{eq:linsys}
A\mathbf{x}=\mathbf{p},
\end{equation}
where the vector $\mathbf{x}\in\R^n$ collects the pixels of the image to be reconstructed, the vector $\mathbf{p}\in\R^m$ gathers the measurements, and the generic entry $a_{ij}$ of the $m\times n$ \emph{projection matrix} $A=[a_{ij}]$ refers to the contribution the $j$-th pixel gives to the $i$-th X-ray. The coefficients $a_{ij}$s can be computed in different ways, according to the employed discrete models. In the grid model, $a_{ij}=1$ when the pixel $j$ belongs to the line $L_i$, while $a_{ij}=0$ otherwise. In the discrete strip model, we have $a_{ij}=1$ when the pixel $j$ belongs to the strip $L_i$, and $a_{ij}=0$ otherwise (see Figure \ref{fig:models}). Note that in a Dirac model, as well as in the grid model, $A$ is a binary matrix. In what follows, for a better visualization of the image, we will identify a pixel $(\xi,\eta)$ with the unit square $[\xi,\xi+1)\times[\eta,\eta+1)$.

Let $S=\{(a_r,b_r): r=1,\ldots,d\}$ be a set of $d$ lattice directions, and $\mathcal{A}=\{(i,j)\in\mathbb{Z}^2\::\:0\leq i<M,\:0\leq j<N\}$ the grid consisting of the pixels of the image, where $MN=n$. The so-called Katz condition states that
\begin{equation*}
\sum_{r=1}^da_r\geq M \quad \text{ or } \quad \sum_{r=1}^d|b_r|\geq N.
\end{equation*}
In this case uniqueness of reconstruction is guaranteed inside the grid $\mathcal{A}$ (\cite{Ka}). Differently, if
\begin{equation}\label{eq:valid}
h=\sum_{r=1}^da_r< M \quad \text{ and } \quad k=\sum_{r=1}^d|b_r|< N,
\end{equation}
then we say that $S$ is a \emph{valid} set of directions for $\mathcal{A}$. For $r=1,\dots,d$ denote
\begin{equation}\label{eq:binomials}
f_{(a_r,b_r)}(x,y)=
\begin{cases}
x^{a_r}y^{b_r}-1&\text{if}\ a_r\neq 0,b_r>0,\\
x^{a_r}-y^{-b_r}&\text{if}\ a_r\neq 0,b_r<0,\\
x-1&\text{if}\ a_r=1,b_r=0,\\
y-1&\text{if}\ a_r=0,b_r=1.
\end{cases}
\end{equation}

\noindent Further, let
\begin{equation}\label{eq:product}
F_S(x,y)=\prod\limits_{r=1}^d f_{(a_r,b_r)}(x,y).
\end{equation}
For any function $g:\mathcal{A}\to\R$, its \emph{generating function} is the polynomial defined by
$$G_g(x,y)=\sum_{(i,j)\in \mathcal{A}}g(i,j)x^{i}y^{j}.$$
A monomial $mx^iy^j\in\Z[x,y]$ can be associated to the lattice point $(i,j)$, together with its \emph{weight} $m$. If $|m|>1$ we say that $(i,j)$ is a \emph{multiple point} and $|m|$ is its \emph{multiplicity}. Therefore, a generating function corresponds geometrically to a lattice set whose points have associated multiplicities. In particular, the \emph{support} of $g$ is the set of lattice points given by $supp(g)=\{(i,j)\in\mathcal{A}:\:g(i,j)\neq 0\}$.

The \emph{line sum}, or \emph{projection}, of $g$ along the lattice line with equation $ay=bx+t$ is defined as $\sum_{aj=bi+t}g(i,j)$. Note that the function $f$, generated by $F_S(x,y)$, has zero line sums along the lines taken in the directions in $S$ (see \cite{Ha-Ti}). Moreover, being $S$ valid for $\mathcal{A}$, $supp(f)$ is contained in $\mathcal{A}$.

For a polynomial $G(x,y)$, we denote by $G^+(x,y)$ (resp., $G^-(x,y)$) the polynomial consisting of the monomials of $G(x,y)$ having positive (resp., negative) coefficients. The sets consisting of the lattice points (counted with their multiplicities) corresponding to $G(x,y), G^+(x,y), G^-(x,y)$ are here denoted by $G$, $G^+$ and $G^-$, respectively.

A function $g:\mathcal{A}\to\R$ is said to be an \emph{$S$-ghost} if it has zero sums along all lines having directions belonging to a given set $S$ of lattice directions. If $supp(g)=\emptyset$, then $g$ is called \emph{trivial ghost}. If $G(x,y)=G_g(x,y)$, then the pair $G=(G^+,G^-)$ is a \emph{(weakly) bad configuration}, and consists of two sets that have the same absolute sums along all lines with directions taken in $S$, up to count each pixel with its proper multiplicity. Consequently, ghosts are responsible of ambiguous outputs in tomographic reconstructions. A \emph{binary} $S$-ghost is an $S$-ghost where $g:\mathcal{A}\to\{-1,0,1\}$. In this case no multiple point belongs to $G=(G^+,G^-)$, which is called \emph{bad configuration}. See also \cite{BDHP,SVCE,SVACHA,SVANOR} for recent results concerning ghosts in discrete tomography.

\begin{rem}
When the tomographic problem is modeled as a linear system $A\mathbf{x}=\mathbf{p}$, ghosts correspond to non-zero solutions of the homogeneous system $A\mathbf{x}=\mathbf{0}$, since these can be added to any solution of \eqref{eq:linsys}, still returning a solution. If $g:\mathcal{A}\to\R$ is an $S$-ghost, then the corresponding solution of $A\mathbf{x}=\mathbf{0}$, still called \emph{ghost}, is denoted by $\mathbf{x}_g$. If we index the points of $\mathcal{A}$ according to some ordering, then the $\mu$-th entry of $\mathbf{x}_g$ is $x_{\mu}=g(i,j)$ if and only if $(i,j)$ is the $\mu$-th pixel.
\end{rem}

The number of entries (also called bins) of the  projection array $\mathbf{p}$ depends both on the projection angles and on the  size of the lattice grid. For a direction $(a,b)$ and an $M\times N$-sized lattice grid, there are $(M-a)\abs{b}+(N-\abs{b})a + a\abs{b}$ bins, so that the size of $\mathbf{p}$ is linear in the grid dimensions and in the number $d$ of employed directions (see \cite{Ha-Ti}).

If the Katz criterion holds, then no ghost exists inside the given grid. On the other side, if the Katz condition is not fulfilled, then uniqueness of reconstruction is not allowed without introducing some extra information, since ghosts always appear. It is worth clarifying that \textit{extra information} means any kind of prior knowledge concerning the tomographic problem, such as that the object to be reconstructed is binary, or that it is contained in a finite grid. For instance, a special class of geometric objects, widely considered in the literature, is represented by \emph{additive} sets (see \cite{FLRS} for further information and related results). Indeed, a finite set $E\subset\Z^2$ is uniquely determined by its $X$-rays in the coordinate directions if and only if $E$ is additive. More generally, the notions of additivity and uniqueness are equivalent when two directions are employed, whereas, for three or more directions, additivity is more demanding than uniqueness (see \cite{FLRS,FS} for details). However, uniqueness results can be achieved even without the additivity assumption (see \cite{BDP3,BDP2014,GLW} and the related bibliographies).

We have therefore two different reconstruction approaches, both with positive aspects and drawbacks. If the Katz limitations hold, then uniqueness is guaranteed, but many short directions (namely, whose entries are small), or few long directions, must be considered. Differently, when the Katz inequalities do not hold, then uniqueness could be obtained by some convenient combination of few short directions and further conditions.

From the above discussion we are led to focus on the problem of reconstructing an unknown image by exploiting sets of directions that guarantee uniqueness in a given lattice grid.

In this paper we provide a solution to this problem in the case $\mathbf{x}\in\{0,1\}^n$, and when suitably selected valid sets of four directions are employed.

\section{A uniqueness result for binary reconstructions}\label{sec:theoretical}
Consider now the linear system $A\mathbf{x}=\mathbf{p}$, $A\in\R^{m\times n},\mathbf{p}\in\R^m$. If $\mathbf{p}$ collects consistent data, then the linear system supports a solution, even if, due to ghosts, usually many outputs are allowed. In this case one can try to find a particular solution $\mathbf{x}^{\ast}$, and then to include in the problem some extra information, in order to modify $\mathbf{x}^{\ast}$ so that the new solution matches the added requirements. In case of BT, the solution $\mathbf{x}^{\ast}$ having minimal Euclidean norm is of special interest. This depends on different reasons. For instance, it can be easily approximated by iterative algorithms, and its theoretical properties are well known from the singular value decomposition $svd(A)$ of the matrix $A$ (see for instance \cite{GV} for details). Also, in \cite{BFHT} it was shown that all binary solutions of $A\mathbf{x}=\mathbf{p}$ have equal distance
\begin{displaymath}
R=\sqrt{\frac{\|\mathbf{p}\|_1}{d}-\|\mathbf{x}^{\ast}\|_2^2}
\end{displaymath}
to $\mathbf{x}^{\ast}$, being $d$ the number of employed directions. This means that $\mathbf{x}^{\ast}$ is the center of a hypersphere of radius $R$ which contains all the binary solutions. Because of this, in what follows we refer to $\mathbf{x}^{\ast}$ as the \emph{central reconstruction} (or \emph{solution}) of the tomographic problem.

Now, let $S=\{u_1,u_2,u_3,u_4=u_1+u_2\pm u_3\}$ be a valid set of four directions for the grid $\mathcal{A}=\{(i,j)\in \Z^2:\: 0\leq i<M,\:0\leq j<N\}$, $\hat{S}=\{(u_1-u_4),(u_2-u_4),(u_1+u_2)\}$, and $D=(\pm S)\cup (\pm\hat{S})$, where $\pm S=\{\pm u_r\mid r=1,\ldots,4\}$ and $\pm\hat{S}=\{\pm(u_1-u_4),\pm(u_2-u_4),\pm(u_1+u_2)\}$. The set $D$, therefore, is not a set of directions, but a set of pairs, since the entries of its elements are not necessarily coprime integers. Define the two disjoint sets $A,B$ as follows:
\begin{eqnarray*}
A&:=&\{(a,b)\in D:\; \abs{a}> \abs{b}\},\\ B&:=&\{(a,b)\in D:\; \abs{b}> \abs{a}\}.
\end{eqnarray*}
Moreover, if $|a|=|b|$ for some $(a,b)\in D$, we then include $(a,b)$ in $A$ if $\min\{M-h,N-k\}=M-h$, while $(a,b)\in B$ otherwise ($h,k$ defined as in \eqref{eq:valid}). Thus we have $D=A\cup B$, where one of the sets $A,B$ may be empty. The following result has been obtained in \cite{BDP1}, and it represents a criterion for preventing the existence of binary $S$-ghosts in $\mathcal{A}$.

\begin{teo}\label{teo:uniqueness}
Let $S=\{u_{1},u_{2},u_{3},u_{4}=u_{1}+u_{2}\pm u_{3}\}$ be a valid set for the lattice grid $\mathcal{A}=\{(i,j)\in \Z^2: \: 0\leq i<M,\:0\leq j<N\}$. Suppose that $g:\mathcal{A}\to \{-1,0,1\}$ has zero line sums along the lines with direction in $S$. Then $g$ is identically zero if and only if
\begin{eqnarray}\label{iff1a}
\min_{(a,b)\,\in\, A} \abs{a}\geq \min \{M-h,N-k\},\\
\label{iff1b}
\min_{(a,b)\,\in\, B} \abs{b} \geq \min \{M-h,N-k\},
\end{eqnarray}
and
\begin{eqnarray}\label{conditionB}
M-h < N-k\,\,\Rightarrow\,\, \forall(a,b)\in B: \,|a|\geq M-h\, \text{ or }\, |b|\geq N-k,
\\\label{conditionA}
N-k < M-h\,\, \Rightarrow\,\, \forall(a,b)\in A: \,|a|\geq M-h\, \text{ or }\, |b|\geq N-k,
\end{eqnarray}
where, if one of the sets $A,B$ is empty, the corresponding condition \eqref{iff1a} or \eqref{iff1b} drops.
\end{teo}

\begin{rem}
A set $S$ of directions satisfying the assumptions of Theorem \ref{teo:uniqueness} always determines a weakly bad configuration $G$ having a double point and prevents $G$ from being modified into a bad configuration still remaining inside the grid, which is the reason that guarantees binary uniqueness. There is no result like Theorem \ref{teo:uniqueness} for two or three directions, since in these cases the corresponding bad configurations never present a double point, as one can easily check. For $d>4$ directions, a characterization of the sets of directions ensuring the presence of a double point is still missing. Therefore, $d=4$ directions is a minimal choice in view of uniqueness.
\end{rem}

Theorem \ref{teo:uniqueness} provides uniqueness conditions that we can exploit in view of a reconstruction algorithm. We remark that the reconstruction problem is known to be NP-hard for more than two directions (see \cite{GGP}). However, for special sets of directions it can become tractable. We will show that this is the case for sets of directions satisfying the previous assumptions.

\begin{defin}\label{def:goodsets}
A set $S$ satisfying all the assumptions of Theorem \ref{teo:uniqueness} is said to be a \emph{ set of binary uniqueness} for $\mathcal{A}$.
\end{defin}

The collection of the sets of binary uniqueness for $\mathcal{A}$ is denoted by $\mathcal{S}(\mathcal{A})$.

A general criterion for the construction of $\mathcal{S}(\mathcal{A})$ is not known. However, there exist sufficient conditions for a set $S$ to be in $\mathcal{S}(\mathcal{A})$, as in the following corollaries (see \cite{BDP1,DuPa}).

\begin{cor}
If $N$ is odd, then projections taken along directions in the set
$$S=\left\{(1,0),(0,1),\left(\frac{N-1}{2},\frac{N-3}{2}\right),\left(\frac{N-3}{2},\frac{N-1}{2}\right)\right\}$$
uniquely reconstructs an $(N\times N)$-sized binary grid.
\end{cor}

\begin{cor}
Let $S=\{u_i=(a_i,b_i),\:i=1,2,3,4\}$ be a set of lattice directions, $a_1=\min_ia_i$, $b_1=\min_ib_i\geq 0$. Suppose that $r_1,r_2,s_1,s_2$ exist such that
\begin{displaymath}
\begin{array}{ll}
a_2=a_1+r_1,&b_2=b_1+s_1,\\
a_3=a_1+r_2,&b_3=b_1+s_2,\\
a_4=a_1+a_2+a_3,&b_4=b_1+b_2+b_3,\\
r_1+r_2\geq\frac{M-7a_1}{2},&s_1+s_2\geq\frac{N-7b_1}{2}.
\end{array}
\end{displaymath}
Then $S\in\mathcal{S}(\mathcal{A})$.
\end{cor}

In what follows we show how to match the central reconstruction $\mathbf{x}^{\ast}$ with Theorem \ref{teo:uniqueness}, in order to reconstruct the guaranteed unique binary solution.

\section{The space of ghosts in a lattice grid}\label{sec:ghosts}
We first prove a result which makes us pay special attention to a specific point of $F_S$.

\begin{prop}\label{lem:axis_ghost}
Let $S=\{u_r=(a_r,b_r) : r=1,\ldots,d\}$ be a set of lattice directions, valid for a lattice grid $\mathcal{A}=\{(i,j)\in \Z^2: \: 0\leq i<M,\:0\leq j<N\}$. Then the (weakly) bad configuration $F_S$ associated to $F_S(x,y)$ intersects the $y$-axis.
\end{prop}

\proof If $b_r\geq 0$ for all $r\in\{1,\ldots,d\}$, then the product of $d$ binomials $f_{(a_r,b_r)}(x,y)$ of the form \eqref{eq:binomials} always includes the constant term $+1$ or $-1$, according to the fact that $d$ is even or odd, respectively. By \eqref{eq:product}, this holds for $F_S(x,y)$, meaning that $F_S$ contains the origin. If $b_r<0$ for some $r$, then the previous product always includes a monomial of the form $\alpha y^j$, for some $j\in\mathbb{N}$, and $\alpha\neq 0$, meaning that $F_S$ contains the lattice point $(0,j)$.\qed

In particular, we denote by $\lambda_0$ the pixel of $F_S$ lying on the $y$-axis.

For $S\in\mathcal{S}(\mathcal{A})$, let $A\mathbf{x}=\mathbf{p}_S$ be the linear system modeling the tomographic problem. First of all, we investigate the structure of the existing (non binary) $S$-ghosts in $\mathcal{A}$. Denote by $\mathcal{G}_S$ the set of all ghosts associated to $S$, namely, the set of solutions of the homogeneous system $A\mathbf{x}=\mathbf{0}$. Therefore, $\mathcal{G}_S$ is a subspace of $\R^n$ isomorphic to $\mathrm{null}(A)$, the null-space of $A$, so that $\dim(\mathcal{G}_S)=n-\mathrm{rank}(A)$.

We are interested in investigating how an $S$-ghost included in the grid $\mathcal{A}$ can cover the different pixels of the grid. First of all, from Definition \ref{def:goodsets}, and from the proof of Theorem \ref{teo:uniqueness} (see \cite{BDP1}), it follows that, for any set $S\in\mathcal{S}(\mathcal{A})$, the four directions in $S$ provide a weakly bad configuration, denoted by $F_S$, and consisting of fifteen pixels $\{\lambda_0,\ldots,\lambda_{14}\}$, where one pixel $\lambda_{\delta}$ is counted twice, and the others have weight $\pm 1$. For any $w_0\in\R$, we denote by $F_S(w_0)$ the weighted weakly bad configuration $w_0F_S$ whose pixel $\lambda_0$ has weight $w_0$. In particular, $F_S(1)=F_S$. The pixels of $F_S$ having weight $+1$ are $\lambda_0$ and the pixels obtained by translating $\lambda_0$ along vectors corresponding to the sum of $2$ or $4$ elements of $S$. The pixels of weight $-1$ come from translations of $\lambda_0$ along vectors corresponding to the sum of $1$ or $3$ elements of $S$. Denote by $I^{+}$ (resp., $I^{-}$) the set of indices $t\neq\delta$ such that $\lambda_t\in F_S$ has weight $+1$ (resp., $-1$).

\begin{defin}\label{def:movingregion}
The \emph{enlarging region} associated to $F_S$ is the rectangle $E=\{(i,j): \:0\leq i\leq M-h-1,\:0\leq j\leq N-k-1\}$. Further, for each $(\xi,\eta)\in\mathcal{A}$, define $E^{+}(\xi,\eta)=\{u\in E: \:(\xi,\eta)=\lambda_i+u,\:i\in I^{+}\}$ and $E^{-}(\xi,\eta)=\{u\in E: \:(\xi,\eta)=\lambda_t+u,\:t\in I^{-}\}$.
\end{defin}

\begin{figure}[htbp]
\centering
\includegraphics[scale=0.3]{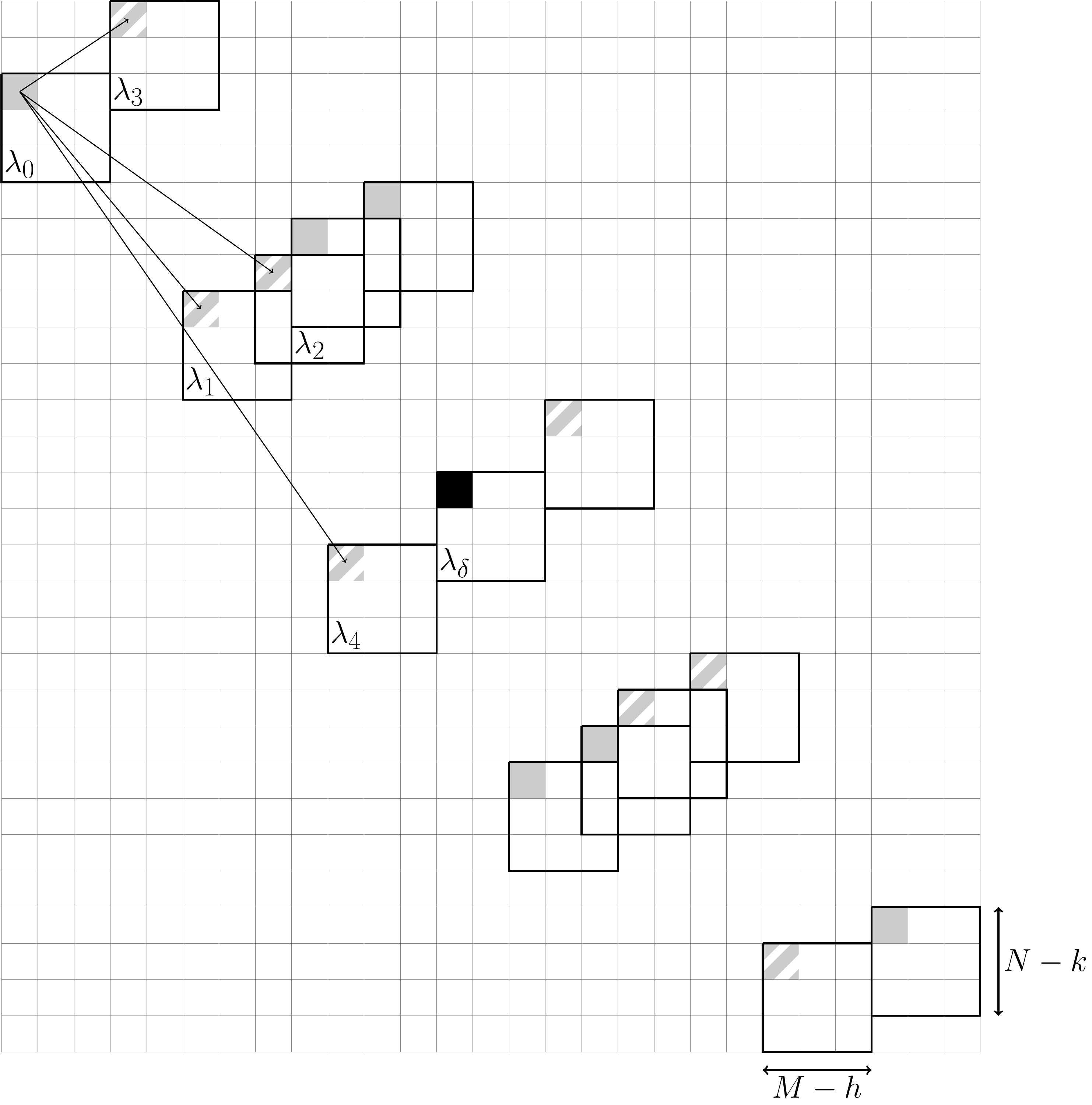}
\caption{The structure of the weakly bad configurations for $M=26$, $N=28$ and $S=\{(5,6),(7,5),(3,-2),(9,13)\}$. In this case $M-h=N-k=3$, so there are nine weakly bad configurations, which overlap. The (fully and striped) colored pixels correspond to $F_S$ and $\lambda_{\delta}$ is the pixels counted twice. Note that the rectangles $\lambda_0+E$ and $\lambda_{\delta}+E$ do not intersect the others.}
\label{fig:overlap}
\end{figure}

The \emph{enlarging region associated to a pixel $\lambda\in F_S$} is the set $\lambda+E$. The collection of the enlarging regions associated to all pixels of $F_S$ is therefore the region where each pixel of $F_S$ can be moved without exceeding the grid sides. Figure \ref{fig:overlap} shows the structure of $F_S$, and of the corresponding enlarging regions, in the case $u_4=u_1+u_2-u_3$ (the other case, $u_4=u_1+u_2+u_3$, leads to a similar configuration). Fully gray colored pixels have weight $w_0$ in $F_S(w_0)$, dashed pixels have weight $-w_0$, and the weight of the black pixel is $2w_0$.

The notion of enlarging region is related to the following theorem (where, as usual, $h,k$ are defined as in \eqref{eq:valid}), which is simply a rephrasing of the results in \cite{Ha-Ti}.

\begin{teo}\label{teo:ghoststructure}
If $S\in\mathcal{S}(\mathcal{A})$, then $\dim(\mathcal{G}_S)=(M-h)(N-k)$ and, for all $g\in\mathcal{G}_S$, $15\leq |supp(g)|\leq 15(M-h)(N-k)$.
\end{teo}

\proof By \cite[Theorem 1]{Ha-Ti} any $S$-ghost $g:\mathcal{A}\to\R$ is a linear combination of $(M-h)(N-k)$ linearly independent switching elements. This means that $\dim(\mathcal{G}_S)=(M-h)(N-k)$. Moreover, by \cite[Corollary 1]{Ha-Ti}, a basis of $\mathcal{G}_S$ can be obtained by considering the $(M-h)(N-k)$ switching elements $F_S(1)+u$ for all $u\in E$, so that $|F_S(1)|\leq |supp(g)|\leq |F_S(1)||E|$. Since $|F_S(1)|=15$, it is $15\leq |supp(g)|\leq 15(M-h)(N-k)$.\qed

Geometrically, $D$ represents the set of vectors along which the double point of $F_S$ has to be translated in order to reach the other points of $F_S$ (see \cite{BDP1}). In the next lemma we prove that the enlarging regions containing $\lambda_0$ and the double pixel do not intersect the others.

\begin{prop}\label{lem:ghost_translation}
Let $\mathcal{A}$ be defined as before, $S\in\mathcal{S}(\mathcal{A})$, $F_S=\{\lambda_0,\ldots,\lambda_{14}\}$ the weakly bad configuration determined
by $S$ in $\mathcal{A}$. Denote by $\lambda_{\delta}$ the point of $F_S$ which is counted twice. Then
\begin{displaymath}
\forall u\in E, t\neq0,\delta: \lambda_t+u\notin (\lambda_0+E)\cup(\lambda_{\delta}+E).
\end{displaymath}
\end{prop}

\proof The fact that $\lambda_0+E$ does not intersect other rectangles comes from the last paragraph of \cite[Section 2]{Ha-Ti}, where it is pointed out that all pixels in $\lambda_0+E$ get value $\pm 1$.

On the other side, consider the double pixel $\lambda_{\delta}$. Let $v=(v_1,v_2)\in D$ and assume that $\min\{M-h,N-k\}=M-h$. If $v\in A$, then, by condition \eqref{iff1a} of Theorem \ref{teo:uniqueness}, it is $|v_1|\geq M-h$, which means that the enlarging region of pixel $\lambda_t=\lambda_{\delta}+v$ cannot intersect $\lambda_{\delta}+E$, since $E$ has horizontal size equal to $M-h$. If $v\in B$, then by condition \eqref{conditionB} of Theorem \ref{teo:uniqueness} we get $|v_1|\geq M-h$ or $|v_2|\geq N-k$. In the first case we reach the same conclusion as above. In the second case the rectangles do not intersect as well, since the vertical size of $E$ is $N-k$.

If $\min\{M-h,N-k\}=N-k$, the proof is similar. Therefore, the enlarging rectangle of the double pixel has no overlaps with other rectangles.\endproof

In particular, since $\lambda_0+E$ does not intersect any other rectangle, then, for any $u\in E$, we have (note that $0\in I^+$)
\begin{equation}\label{eq:singlecoefficient}
\begin{aligned}
E^{+}(\lambda_0+u)&=\{v\in E: \:\lambda_0+u=\lambda_i+v,\:i\in I^{+}\}=\{u\},\\
E^{-}(\lambda_0+u)&=\{v\in E: \:\lambda_0+u=\lambda_t+v,\:t\in I^{-}\}=\emptyset.
\end{aligned}
\end{equation}

\begin{rem}\label{rem:numberofintersection}
Differently from what stated in Lemma \ref{lem:ghost_translation}, in case $i,t\neq 0,\delta$ it could be $(\lambda_i+E)\cap(\lambda_t+E)\neq\emptyset$.
\end{rem}

\begin{ex}\label{ex:ghost size}
Let $\mathcal{A}$ be a square lattice grid of size $51$ and consider the set of directions
$$S=\{u_1=(3,5),u_2=(5,3),u_3=(16,15),u_4=(24,23)\}.$$
The fourth direction is the sum of the previous three and it can be easily checked that all the assumptions of Theorem \ref{teo:uniqueness} hold, and consequently $S\in\mathcal{S}(\mathcal{A})$. The polynomial associated to $F_S$ is
$$\begin{array}{lll}F_S(x,y)&=&x^{48}y^{46} - x^{45}y^{41} - x^{43}y^{43} + x^{40}y^{38} - x^{32}y^{31} + x^{29}y^{26} + x^{27}y^{28} \\ &&- 2x^{24}y^{23} + x^{21}y^{18} + x^{19}y^{20} - x^{16}y^{15} + x^8y^8 - x^5y^3 - x^3y^5 + 1.
\end{array}$$
Moreover, it is $M-h=3$ and $N-k=5$, so that $E=\{(\xi,\eta):0\leq\xi\leq 2,0\leq\eta\leq 4\}$. By Theorem \ref{teo:ghoststructure}, the support of any $S$-ghost $g$ is $supp(g)=\{\lambda_i+E:0\leq i\leq14\}$, where $\lambda_i=(\alpha_i,\beta_i)$, being $x^{\alpha_i}y^{\beta_i}$ anyone of the $15$ monomials of $F_S(x,y)$. In particular, $\lambda_0=(0,0)$ and $\lambda_{\delta}=u_4=u_1+u_2+u_3=(24,23)$. Note that, according to Lemma \ref{lem:ghost_translation}, the rectangles $\lambda_{\delta}+E$ and $\lambda_0+E$ are always disjoint from the others, as one can easily check by considering the exponents of the monomials of $F(x,y)$. However, if $i,t\neq 0,\delta$ possible overlaps might occur between $\lambda_i+E$ and $\lambda_t+E$ (see Remark \ref{rem:numberofintersection}). For instance, if $\lambda_1=(3,5)$ and $\lambda_2=(5,3)$, we have $\lambda_1+E=\{(\xi,\eta)\in\mathcal{A} :\:3\leq\xi\leq 5, 5\leq\eta\leq 9\}$, and $\lambda_2+E=\{(\xi,\eta)\in\mathcal{A} :\:5\leq\xi\leq 7, 3\leq\eta\leq 7\}$, so that all pixels $(5,\eta)$ such that $5\leq \eta\leq 7$ belong to both sets.
\end{ex}

\section{Binary reconstruction from the central solution}\label{sec:reconstruction}
We can exploit Theorem \ref{teo:uniqueness} to select sets $S$ of four directions leading to linear systems of equations that admit only one binary solution $\mathbf{\overline{\mathbf{x}}}$. In this case Theorem \ref{teo:ghoststructure} provides lower and upper bounds on the size of wrongly reconstructed pixels when $\mathbf{\overline{\mathbf{x}}}$ is approximated by a generic solution of \eqref{eq:linsys}. Following \cite{BFHT}, we are induced to focus on the binary rounding $\mathbf{\overline{\mathbf{x}}}^{\ast}$ of the central solution, and, by Theorem \ref{teo:ghoststructure}, to work just in the region possibly covered by ghosts. The following results lead to the exact reconstruction of a binary image from the binary rounding of the central solution.

Let $F_S=\{\lambda_0,\ldots,\lambda_{14}\}$ be the weakly bad configuration associated to a set $S=\{u_1,u_2,u_3,u_4=u_1+u_2\pm u_3\}\in\mathcal{S}(\mathcal{A})$. For $u=(p,q)\in E$, let $G_u=F_S+u$, and let $g_u:\mathcal{A}\to\R$ be the $S$-ghost generated by $x^py^qF_S(x,y)$. In case $u_1+u_2=u_3+u_4$, it is $\lambda_{\delta}=u_1+u_2=u_3+u_4$ (see Figure \ref{fig:overlap}), so that $I^+$ contains $6$ elements and $I^-$ contains $8$ indices. It results
\begin{equation}\label{eq:base_ghost}
g_u(\xi,\eta)=\left\{\begin{array}{rl} 0 &\text{if $(\xi,\eta)\notin G_u$}\\
1&\text{if $(\xi,\eta)=\lambda_i+u$, $i\in I^{+}$}\\
-1 &\text{if $(\xi,\eta)=\lambda_i+u$, $i\in I^{-}$}\\
2 &\text{if $(\xi,\eta)=\lambda_{\delta}+u$}.
\end{array}
\right.
\end{equation}
In case $u_1+u_2+u_3=u_4$ we have an analogous definition, just observing that $\lambda_{\delta}=u_4=u_1+u_2+u_3$ has weight $-2$, and changing the sets $I^{+},I^{-}$ accordingly. If $\mathbf{y}$ is any solution of $A\mathbf{x}=\mathbf{p}_S$, then
\begin{displaymath}
\mathbf{y}(\xi,\eta)=\overline{\mathbf{x}}(\xi,\eta)+\sum_{u\in E}\alpha_ug_u(\xi,\eta),
\end{displaymath}
for all $(\xi,\eta)\in\mathcal{A}$, and for suitable coefficients $\alpha_u\in\R$. Let $\{\alpha^{\ast}_{u}\in\R: \:u\in E\}$ be the set of real values corresponding to the minimal norm solution. In particular, for all $(\xi,\eta)\in\mathcal{A}$, the minimal norm solution is
\begin{displaymath}
\mathbf{\mathbf{x^{\ast}}}(\xi,\eta)=\overline{\mathbf{x}}(\xi,\eta)+\sum_{u\in E}\alpha^{\ast}_{u}g_u(\xi,\eta).
\end{displaymath}
For $(\xi,\eta)\in\mathcal{A}$, we call \emph{minimal weight of $(\xi,\eta)$} the weight $w^{\ast}(\xi,\eta)$ given to $(\xi,\eta)$ by the minimum norm solution. By Lemma \ref{lem:ghost_translation} and Remark \ref{rem:numberofintersection}, it results (see also Definition \ref{def:movingregion})
\begin{displaymath}
w^{\ast}(\xi,\eta)=\sum_{u\in E}\alpha^{\ast}_{u}g_u(\xi,\eta)=
\left\{\begin{array}{ll}
0 &\text{if }(\xi,\eta)\notin H,\:\:H=\bigcup_{u\in E}G_u,\\
2\alpha^{\ast}_u &\text{if $(\xi,\eta)=\lambda_{\delta}+u$},\\
\displaystyle{\sum_{u\in E^{+}(\xi,\eta)}\alpha^{\ast}_{u}-\sum_{u\in E^{-}(\xi,\eta)}\alpha^{\ast}_{u}}&\text{otherwise}.
\end{array} \right.
\end{displaymath}
Therefore, $\mathbf{x^{\ast}}(\xi,\eta)=\overline{\mathbf{x}}(\xi,\eta)+w^{\ast}(\xi,\eta)$, so that
\begin{equation}\label{eq:roundingcases}
\overline{\mathbf{x}}(\xi,\eta)=\mathbf{x^{\ast}}(\xi,\eta)-w^{\ast}(\xi,\eta)=\left\{\begin{array}{ll}
\mathbf{x^{\ast}}(\xi,\eta)&\text{if\:}(\xi,\eta)\notin H,\\\\
\mathbf{x^{\ast}}(\xi,\eta)-2\alpha^{\ast}_{u},&\text{if\:}(\xi,\eta)=\lambda_{\delta}+u,\: u\in E,\\
\displaystyle{\mathbf{x^{\ast}}(\xi,\eta)-\sum_{u\in E^{+}(\xi,\eta)}\alpha^{\ast}_{u}+\sum_{u\in E^{-}(\xi,\eta)}\alpha^{\ast}_{u}}&\text{otherwise}.
\end{array}\right.
\end{equation}

Therefore, $\overline{\mathbf{x}}(\xi,\eta)$ can be reconstructed from $\mathbf{x}^{\ast}(\xi,\eta)$ once we can compute explicitly $w^{\ast}(\xi,\eta)$ for all $(\xi,\eta)\in\mathcal{A}$.

The following theorem proves that the coefficients of the weakly bad configurations can be computed from the values the central solution $\mathbf{x^{\ast}}$ takes in the pixels of the enlarging region of $\lambda_0$. Denote by $\mathrm{round}(\gamma)$ the integer rounding of a real number $\gamma\in\R$.

\begin{teo}\label{teo:rounding}
Let $\mathcal{A}=\{(\xi,\eta)\in \Z^2: \: 0\leq \xi<M,\:0\leq \eta<N\}$, $S\in\mathcal{S}(\mathcal{A})$, and let $\mathbf{x^{\ast}}$ be the central solution of $A\mathbf{x}=\mathbf{p}_S$. Then, for all $u\in E$ it results
\begin{equation}\label{eq:minimalweightcomputation}
\alpha^{\ast}_{u}=\mathbf{x^{\ast}}(\lambda_0+u)-\mathrm{round}(\mathbf{x^{\ast}}(\lambda_0+u)).
\end{equation}
\end{teo}

\proof We give the proof when $\mathbf{p}_S$ consists of projections along directions belonging to a set $S$ such that $u_1+u_2=u_3+u_4$, and calling $\lambda_{\delta}$ the double pixel (as remarked above, the other case where $u_1+u_2+u_3=u_4$  follows similarly once we change $\lambda_{\delta}$ and the sets $I^{+},I^{-}$ accordingly). For each $u\in E$, let $f:\R\rightarrow\R$ be the following function:
\begin{displaymath}
f(\alpha_u)=\sum_{i\in I^+}\left(\overline{\mathbf{x}}(\lambda_i+u)+\alpha_u\right)^2+\sum_{i\in I^-}\left(\overline{\mathbf{x}}(\lambda_i+u)-\alpha_u\right)^2+
\left(\overline{\mathbf{x}}(\lambda_{\delta}+u)+2\alpha_u\right)^2.
\end{displaymath}
Let $\mathbf{y}$ be a real-valued solution of $A\mathbf{x}=\mathbf{p}_S$. By assuming $H=\bigcup_{u\in E}G_u$, we get
\begin{eqnarray*}
\|\mathbf{y}\|_2^2&=&\sum_{(\xi,\eta)\in\mathcal{A}}\left(\overline{\mathbf{x}}(\xi,\eta)+\sum_{u\in E}\alpha_ug_u(\xi,\eta)\right)^2\\
&=&\sum_{(\xi,\eta)\notin H}\overline{\mathbf{x}}^2(\xi,\eta)+\sum_{(\xi,\eta)\in H}\left(\overline{\mathbf{x}}(\xi,\eta)+\sum_{u\in E}\alpha_ug_u(\xi,\eta)\right)^2\\
&=&\sum_{(\xi,\eta)\notin H}\overline{\mathbf{x}}^2(\xi,\eta)+\sum_{u\in E}\left[\sum_{i\in I^+}(\overline{\mathbf{x}}(\lambda_i+u)+\alpha_u)^2+
\sum_{i\in I^-}(\overline{\mathbf{x}}(\lambda_i+u)-\alpha_u)^2+(\overline{\mathbf{x}}(\lambda_{\delta}+u)+2\alpha_u)^2\right]\\
&=&\sum_{(\xi,\eta)\notin H}\overline{\mathbf{x}}^2(\xi,\eta)+\sum_{u\in E}f(\alpha_u).
\end{eqnarray*}
The central solution $\mathbf{x^{\ast}}$ is obtained when $\|\mathbf{y}\|_2^2$ attains its minimum value. Note that $f(\alpha_u)\geq 0$ for all $u\in E$. Therefore $\|\mathbf{y}\|_2^2$ is the sum of the constant term $\sum_{(\xi,\eta)\notin H}\overline{\mathbf{x}}^2(\xi,\eta)$, and of $|E|$ copies of the non-negative function $f$ applied on one variable $\alpha_u$, for all $u\in E$. Consequently, the minimum of $\|\mathbf{y}\|_2^2$ is obtained by minimizing $f$, separately with respect to each variable. Computing the derivative we get
\begin{eqnarray*}
f'(\alpha_u)&=&2\left(\sum_{i\in I^+}\left(\overline{\mathbf{x}}(\lambda_i+u)+\alpha_u\right)-\sum_{i\in I^-} \left(\overline{\mathbf{x}}(\lambda_i+u)-\alpha_u\right)+ 2\left(\overline{\mathbf{x}}(\lambda_{\delta}+u)+2\alpha_u\right)\right)\\
&=&2\left(\sum_{i\in I^+}\overline{\mathbf{x}}(\lambda_i+u)-\sum_{i\in I^-} \overline{\mathbf{x}}(\lambda_i+u)+2\overline{\mathbf{x}}(\lambda_{\delta}+u)+18\alpha_u\right),
\end{eqnarray*}
being $|I^+|=6$ and $|I^-|=8$. Therefore, the minimum of $f$ is obtained when
\begin{displaymath}
\alpha_{u,\min}=\alpha^{\ast}_{u}=\frac{\sum_{i\in I^-}\overline{\mathbf{x}}(\lambda_i+u)-\sum_{i\in I^+}\overline{\mathbf{x}}(\lambda_i+u)-2\overline{\mathbf{x}}(\lambda_{\delta}+u)}{18}.
\end{displaymath}
Note that for all $u\in E$ it results
\begin{equation}\label{eq:bounds}
-\frac{4}{9}\leq\alpha^{\ast}_{u}\leq\frac{4}{9},
\end{equation}
where the lower bound is attained if $\overline{\mathbf{x}}(\lambda_i)=0$ for all $i\in I^-$, $\overline{\mathbf{x}}(\lambda_i)=1$ for all $i\in I^+$ and $\overline{\mathbf{x}}(\lambda_{\delta})=1$, while the upper bound is attained if $\overline{\mathbf{x}}(\lambda_i)=0$ for all $i\in I^+$,  $\overline{\mathbf{x}}(\lambda_{\delta})=0$, and $\overline{\mathbf{x}}(\lambda_i)=1$  for all $i\in I^-$.

By Lemma \ref{lem:ghost_translation}, each pixel in $\lambda_0+E$ does not belong to $\lambda_i+E$ for $i\neq 0$. For all $u\in E$ this implies (see Equation \eqref{eq:singlecoefficient}) that there is only one coefficient for each pixel in $\lambda_0+E$. Since $\alpha^{\ast}_{u}\in \left[-\frac{4}{9},\frac{4}{9}\right]$, then we get
\begin{displaymath}
\mathrm{round}(\mathbf{x^{\ast}}(\lambda_0+u))= \mathrm{round}\left(\overline{\mathbf{x}}(\lambda_0+u)+\alpha^{\ast}_{u}\right)=\overline{\mathbf{x}}(\lambda_0+u).
\end{displaymath}
Therefore, the binary solution $\overline{\mathbf{\mathbf{x}}}$ is exactly reconstructed in $\lambda_0+E$. This also allows to compute explicitly the value of each $\alpha^{\ast}_{u}$, namely:
\begin{displaymath}
\mathbf{x^{\ast}}(\lambda_0+u)=\overline{\mathbf{x}}(\lambda_0+u)+\alpha^{\ast}_{u}\:\:\:\:\Rightarrow\:\:\:\:
\alpha^{\ast}_{u}=\mathbf{x^{\ast}}(\lambda_0+u)-\overline{\mathbf{x}}(\lambda_0+u)=\mathbf{x^{\ast}}(\lambda_0+u)- \mathrm{round}(\mathbf{x^{\ast}}(\lambda_0+u))
\end{displaymath}
and the theorem is proven.\qed

\begin{cor}\label{cor:roundingreconstruction}
Let $\mathcal{A}$ be a grid defined as before, $S\in\mathcal{S}(\mathcal{A})$, $\mathbf{\mathbf{x^{\ast}}}$ be the central solution of $A\mathbf{x}=\mathbf{p}_S$. Then the unique binary solution $\overline{\mathbf{x}}$ is uniquely and explicitly reconstructible from $\mathbf{\mathbf{x^{\ast}}}$.
\end{cor}

\proof From the previous theorem, the values $\alpha_u^*$ are determined. Recalling that $w^*=\sum_{u\in E}\alpha^{\ast}_{u}g_u$, then Equation \eqref{eq:roundingcases} allows us to retrieve all pixel values of $\mathbf{x^*}$.\qed

\begin{rem}\label{rem:ambiguous}
Thanks to \eqref{eq:bounds}, a set $S\in\mathcal{S}(\mathcal{A})$ guarantees that $\overline{\mathbf{x}}$ can be exactly reconstructed from $\mathbf{\mathbf{x^{\ast}}}$. This implies that no entry of $\mathbf{\mathbf{x^{\ast}}}$ gets value $\frac{1}{2}$, which is the case leading to ambiguities in \cite[Corollary 6]{BFHT}.
\end{rem}

\begin{cor}\label{cor:simple_rounding}
If $|E^{+}(\xi,\eta)\cup E^{-}(\xi,\eta)|=1$ for all $(\xi,\eta)\in H=\bigcup_{u\in E}G_u$, then $\mathbf{\overline{x}}(\xi,\eta)=\mathrm{round}(\mathbf{x^{\ast}}(\xi,\eta))$ for all $(\xi,\eta)\in\mathcal{A}$.
\end{cor}

\proof  If $(\xi,\eta)\notin H$ then by \eqref{eq:roundingcases} we have $\mathbf{\overline{x}}(\xi,\eta)=\mathrm{round}(\mathbf{x^{\ast}}(\xi,\eta))$. If $(\xi,\eta)\in H$, since $E^{+}(\xi,\eta)\cup E^{-}(\xi,\eta)$ contains just one element, then there exists just one $\alpha^{\ast}_u\neq 0$, so that, by \eqref{eq:roundingcases}, we get $\overline{\mathbf{x}}(\xi,\eta)=\mathbf{x^{\ast}}(\xi,\eta)\pm\alpha^{\ast}_u$, where the sign of $\alpha^{\ast}_u$ is determined by the index $i$ of the pixel $\lambda_i\in F_S$ whose enlarging region contains $(\xi,\eta)$. Therefore, by \eqref{eq:bounds}, we have
$\mathrm{round}(\mathbf{x^{\ast}}(\xi,\eta))=\mathrm{round}\left(\overline{\mathbf{x}}(\xi,\eta) \pm\alpha^{\ast}_{u}\right)=\overline{\mathbf{x}}(\xi,\eta).$\qed

\subsection{A binary reconstruction algorithm}\label{subsec:BRA_reconstructions}
The reconstruction steps provided by Theorem \ref{teo:rounding} lead to Algorithm \ref{alg:BRA}, called Binary Reconstruction Algorithm (BRA).

\begin{algorithm}[htbp]\label{alg:BRA}
\KwData{A lattice grid $\mathcal{A}$.}
\KwData{$S\in\mathcal{S}(\mathcal{A})$.}
\KwData{A projection vector $\mathbf{p}_S$ along lattice lines having direction in $S$.}
\KwData{An integer number $\kappa$.}
\KwResult{Reconstruction of $\overline{\mathbf{x}}_{\kappa}$, the approximation of $\overline{\mathbf{x}}$ after $\kappa$ iterations.}
\Begin{
\nl Compute the projection matrix $A$ associated to $S$ (see Section \ref{sec:preliminaries}).\\
\nl\label{step:computexast}Compute an approximation of the minimum norm solution $\mathbf{x}_{\kappa}^{\ast}$ of the linear system $A\mathbf{x}=\mathbf{p}_S$, for $\kappa$ iterations.\\
\nl\label{step:weakly bad configuration}Compute the weakly bad configuration $F_S$ associated to $S$ (see Equation \eqref{eq:product}).\\
\nl Compute the starting pixel $\lambda_0$ of $F_S$ (see Lemma \ref{lem:axis_ghost}).\\
\nl\label{step:enlarging region}Compute the enlarging region $E$ (see Definition \ref{def:movingregion}).\\
\nl Compute $\mathrm{round}(\mathbf{x}_{\kappa}^{\ast}(\lambda_0+u))$ for all $u\in E$.\\
\nl\label{step:weights}Compute the weights $\alpha^{\ast}_{u}=\mathbf{x^{\ast}_{\kappa}}(\lambda_0+u)-\mathrm{round}(\mathbf{x^{\ast}_{\kappa}}(\lambda_0+u))$ for all $u\in E$ (see Equation \eqref{eq:minimalweightcomputation}).\\
\nl Compute $E^-(\xi,\eta), E^+(\xi,\eta)$ for all $(\xi,\eta)\in H$ (see Definition \ref{def:movingregion} and Equation \eqref{eq:roundingcases}).\\
\nl\label{step:rounding}Compute $\overline{\mathbf{x}}_{\kappa}(\xi,\eta)=\mathbf{x_{\kappa}^{\ast}}(\xi,\eta)-w_{\kappa}^{\ast}(\xi,\eta)$ for all $(\xi,\eta)\in\mathcal{A}$ by means of Equation \eqref{eq:roundingcases}.\\
\nl \label{step:approximation}Binary round off of the entries of $\overline{\mathbf{x}}_{\kappa}$.\\
\Return $\overline{\mathbf{x}}_{\kappa}$.
}
\caption{BRA.}
\end{algorithm}

The input parameter $\kappa$ relates to the number of required runs of some iterative algorithm that, at Step \ref{step:computexast}, returns a suitable numerical approximation $\mathbf{x}_{\kappa}^{\ast}$ of the minimum norm solution $\mathbf{x}^{\ast}$. In particular, we have always employed the conjugate gradient least squares (CGLS) algorithm, which reveals to be particularly efficient. The last round off at Step \ref{step:approximation} is required in order to ensure that a binary solution is always returned from the numerical approximation of $\mathbf{x}^{\ast}$. Note that such a rounding step ensures that the resulting $\overline{\mathbf{x}}_{\kappa}$ equals the unique existing binary solution even for small $\kappa$, which however depends on the structure and the complexity of the image to be reconstructed (see also Section \ref{sec:numericalapplications}). Indeed, exact reconstruction occurs whenever the value assigned by BRA to each pixel definitely stabilizes on a value different from $\frac{1}{2}$ (see Remark \ref{rem:ambiguous}), both in $H$ and in $\mathcal{A}\setminus H$. In particular, the computation in the region $\mathcal{A}\setminus H$ only depends on CGLS, so, when it returns values close to $\frac{1}{2}$, some extra iterations could be required in order to stabilize the result. Therefore, at some intermediate step, we can also expect local oscillations of $\mathbf{x}_{\kappa}^{\ast}(\xi,\eta)$ around the value $\frac{1}{2}$ for some pixel $(\xi,\eta)\in\mathcal{A}\setminus H$, with the consequent alternative approximation to $0$ or $1$ of the corresponding rounding (Step \ref{step:approximation} of BRA). However, after a suitable number of iterations the process must reach the exact solution, so that the phenomenon disappears, and the convergence stabilizes.

Concerning the complexity of BRA we can argue as follows. In a lattice grid of size $M\times N$, we have $(M-a)|b|+(N-|b|)a+a|b|$ projections in a given direction $(a,b)$. The sum over all directions $(a,b)$ of the number of projections provides the number $m$ of rows of the projection matrix $A$. The number of columns of $A$ is equal to the number $n$ of pixels to be reconstructed, namely, $n=MN$. We first note that the most expensive part is the running of CGLS, which sensitively depends on the number of iterations and the sparsity of the matrix. An empirical estimate (see Section \ref{sec:numericalapplications}) of the needed iterations to exactly reconstruct an image is the length of its side if the grid is squared. If $M\neq N$, we can write $O(\sqrt{MN})$. Since every pixel lies on just one line for each direction in $S$, then every column of $A$ has exactly four nonzero entries, meaning that the matrix $A$ is very sparse (see also Section \ref{subsec:smallexample} for an explicit computation of $A$). This implies that each iteration of CGLS has a cost comparable to $m$, which is generally bigger than $n=NM$. The part of BRA regarding the update of the weights can be estimated by observing that, for each $(\xi,\eta)\in H$, we must determine the corresponding regions $E^-(\xi,\eta)$ and $E^+(\xi,\eta)$. To this, for all $u\in E$ we can compute first of all the possible differences $(\xi,\eta)-u$, which costs $\mathrm{size}(E)\times\mathrm{size}(H)$. Then we must check if these differences provide some $\lambda_i\in F_S$, which requires $O(1)$ since $|F_S|=15$ is fixed. Therefore, the weights' updating costs $\mathrm{size}(E)\times\mathrm{size}(H)$ that is $(M-h)\times (N-k)\times(MN)$.

In conclusion, an estimate of the computational complexity of BRA is $O(\max\{m\sqrt{MN},(M-h)\times (N-k)\times(MN)\})$.

\subsection{A small example}\label{subsec:smallexample}
Let $\overline{\mathbf{x}}$ be the following $5\times 5$ binary image
$$\overline{\mathbf{x}}=\left[\begin{array}{ccccc}
0&1&1&1&1\\
0&1&1&1&1\\
0&0&1&1&0\\
0&0&0&0&0\\
0&0&0&0&0
\end{array}\right].$$
We exploit such a small image in order to explicitly show the steps of BRA. The input data consist of a set $S$ of valid directions for $\mathcal{A}$, a projection vector $\mathbf{p}_S$ and a number $\kappa$ of iterations. We take $\mathcal{A}=\{(i,j)\in \mathbb{Z}^2 :\: 0\leq i<5,\:0\leq j<5\}$, and $S=\{(1,0),(1,2),(0,1),(2,1)\}$. The set $S$ is of the form $S=\{u_1,u_2,u_3,u_4=u_1+u_2-u_3\}$ and satisfies all the assumptions of Theorem \ref{teo:uniqueness} in the lattice grid $\mathcal{A}$, so $S\in\mathcal{S}(\mathcal{A})$. As explained in Section \ref{subsec:BRA_reconstructions}, the choice of $\kappa$ relates to the degree of approximation of the minimum norm solution $\mathbf{x}^{\ast}$. Due to the small size of the phantom, we expect exact reconstruction within very few iterations, so we fix $\kappa=2$.

The projection vector $\mathbf{p}_S$ is
$$\mathbf{p}_S=[2,3,3,2,0,1,1,2,2,1,2,1,0,0,0,0,0,0,4,4,2,0,0,1,1,1,1,2,1,2,1,0,0,0,0,0]^t,$$
consisting of $36$ projections. The first five entries, $2,3,3,2,0$, correspond to the vertical projections (from right to left), the following $1,1,2,2,1,2,1,0,0,0,0,0,0$ are the projections in direction $(2,1)$, the entries $4,4,2,0,0$ give the horizontal projections (from top to bottom), and $1,1,1,1,2,1,2,1,0,0,0,0,0$ are the projections along direction $(1,2)$.

First of all BRA computes the projection matrix $A$ associated to the set $S$ in the grid model (see Section \ref{sec:preliminaries}, and also \cite{HK1}). The central solution $\mathbf{x}^{\ast}$ is approximated by using $\kappa=2$ iterations of the CGLS algorithm so returning its numerical approximation $\mathbf{x}_2^{\ast}$, and we get

$$\mathbf{x}_2^{\ast}=\left[\begin{array}{ccccc}
0.2001&1.0044&1.1276&0.8812&0.8075\\
0.2892&0.9208&0.8217&1.0044&0.9010\\
-0.1200&0.0967&0.6688&0.8415&0.3332\\
-0.2872&-0.1200&0.1363&0.1363&0.0967\\
-0.2575&-0.0408&0.0032&0.2595&0.0670
\end{array}\right].$$

\noindent At Step \ref{step:weakly bad configuration} BRA computes the weakly bad configuration $F_S$ associated to the set $S$, namely, the set of pixels $\lambda_i=(\alpha_i,\beta_i)$, $i\in\{0,\ldots,14\}$ such that $x^{\alpha_i}y^{\beta_i}$ is a term (with coefficient $\pm 1$ or $2$) of the polynomial
\begin{eqnarray*}
F_S(x,y)=\prod\limits_{(a,b)\in S}f_{(a,b)}(x,y)&=&x^4y^4 - x^4y^3 - x^3y^4 + x^3y^3 - x^3y^2 + x^3y - x^2y^3 + 2x^2y^2\\
&&- x^2y + xy^3 - xy^2 + xy - x - y + 1.
\end{eqnarray*}
Since $b_r\geq 0$ for all $r$, the proof of Lemma \ref{lem:axis_ghost} provides $\lambda_0=(0,0)$. In this small example the enlarging region $E$ computed at Step \ref{step:enlarging region} reduces just to the vector $(0,0)$. This means that $F_S$ cannot be moved inside $\mathcal{A}$, and consequently Corollary \ref{cor:simple_rounding} holds, so that the reconstruction can be simply obtained by rounding off the minimum norm solution returned by CGLS after a suitable number of iterations. Note that just two iterations suffice in this case.

\subsection{Applications of BRA}\label{sec:numericalapplications}
\begin{figure}[htbp]
\centering
\includegraphics[scale=0.6, viewport=57 360 502 769, clip=true]{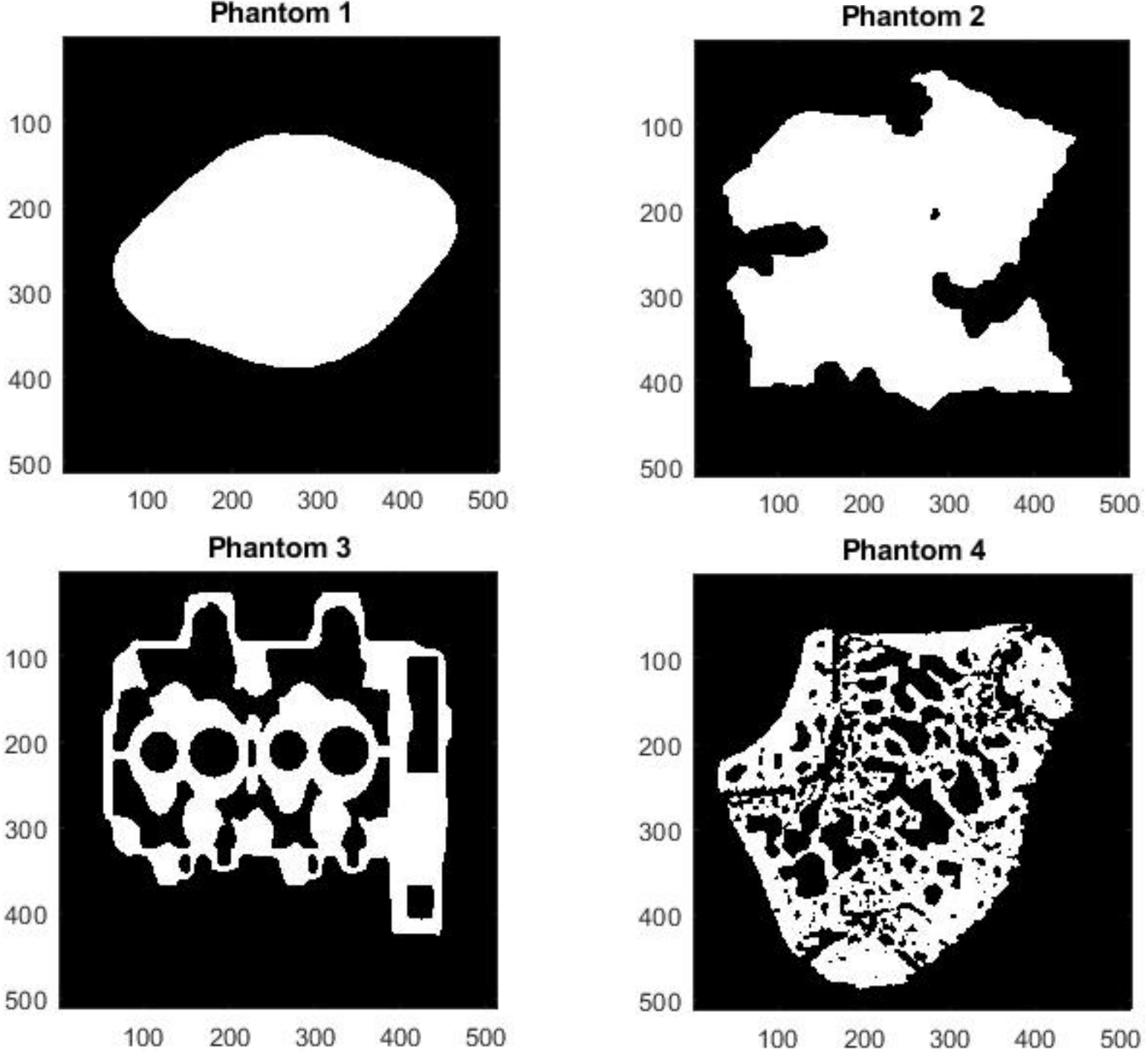}
\caption{Four binary phantoms to which BRA has been applied.}
\label{fig:phantoms}
\end{figure}

We give now a few numerical examples concerning the application of BRA to the reconstruction of binary images in the grid model. We have considered the four $(512\times 512)$-sized binary phantoms presented in \cite{BFHT} (see Figure \ref{fig:phantoms}), and the set of directions $S=\{(80,77),(81,91),(80,83),(241,251)\}$. It is easy to see that $S$ is a set of four valid directions for a $(512\times 512)$-sized grid. Also, $S$ satisfies all the assumptions of Theorem \ref{teo:uniqueness}. We have $M=N=512$, $h=482$, $k=502$, so that $\min\{M-h,N-k\}=M-h=30$ and the set $D$ is
$$D=\{\pm(80,77), \pm(81,91), \pm(80,83), \pm(241,251), \pm(161,174), \pm(160,160), \pm(161,168)\}.$$
Therefore $A=D$, $B=\emptyset$ and conditions \eqref{iff1a} and \eqref{conditionB} hold.

\begin{figure}[htbp]
\centering
\includegraphics[scale=0.6, viewport=56 420 519 772, clip=true]{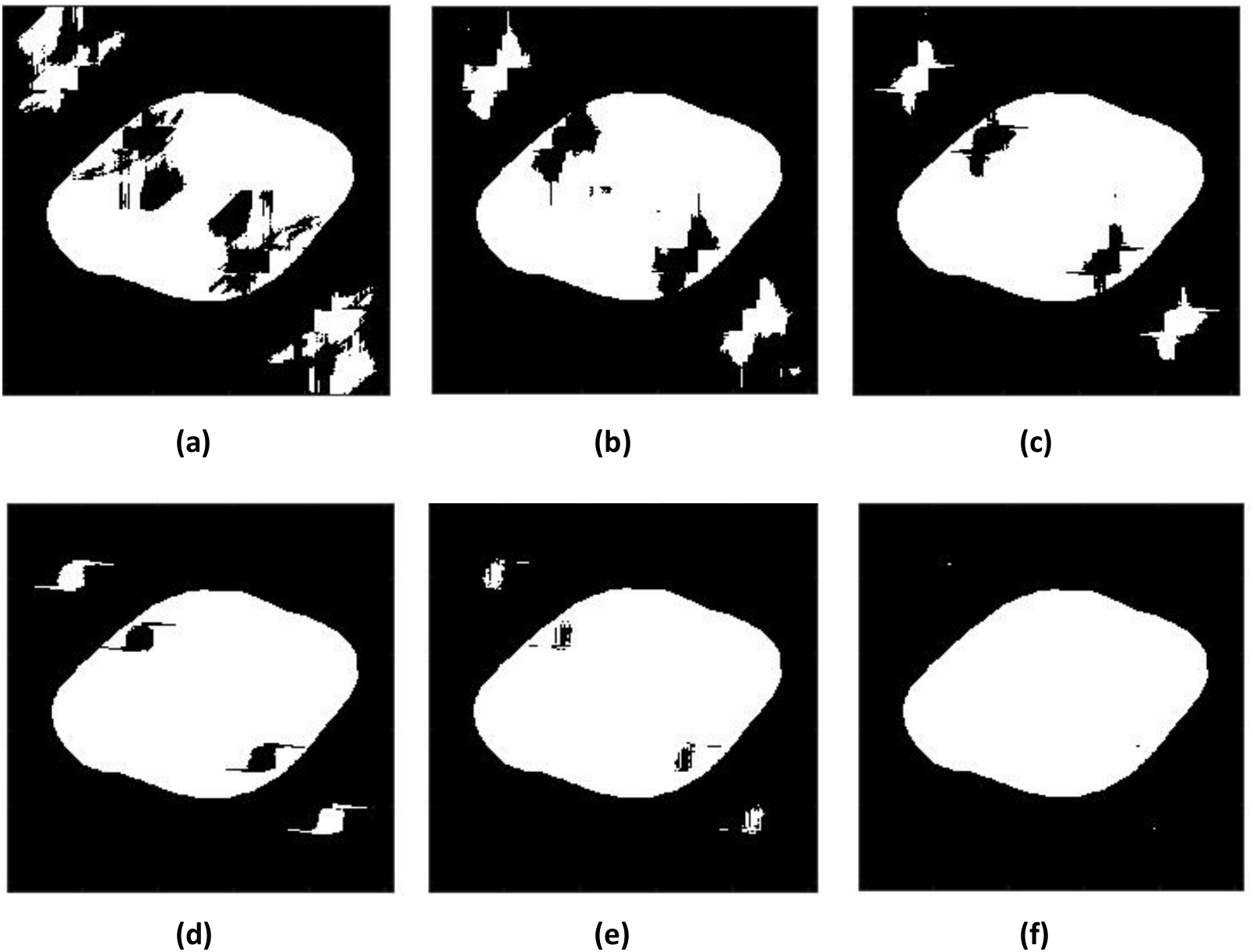}
\caption{Reconstruction of Phantom 1 by BRA for different numbers of iterations: (a) 10 iterations. (b) 50 iterations. (c) 100 iterations. (d) 150 iterations. (e) 200 iterations. (f) 250 iterations. Exact reconstruction is obtained within 350 iterations.}
\label{fig:phantom1}
\end{figure}

\begin{figure}[htbp]
\centering
\includegraphics[scale=0.6, viewport=56 410 519 772, clip=true]{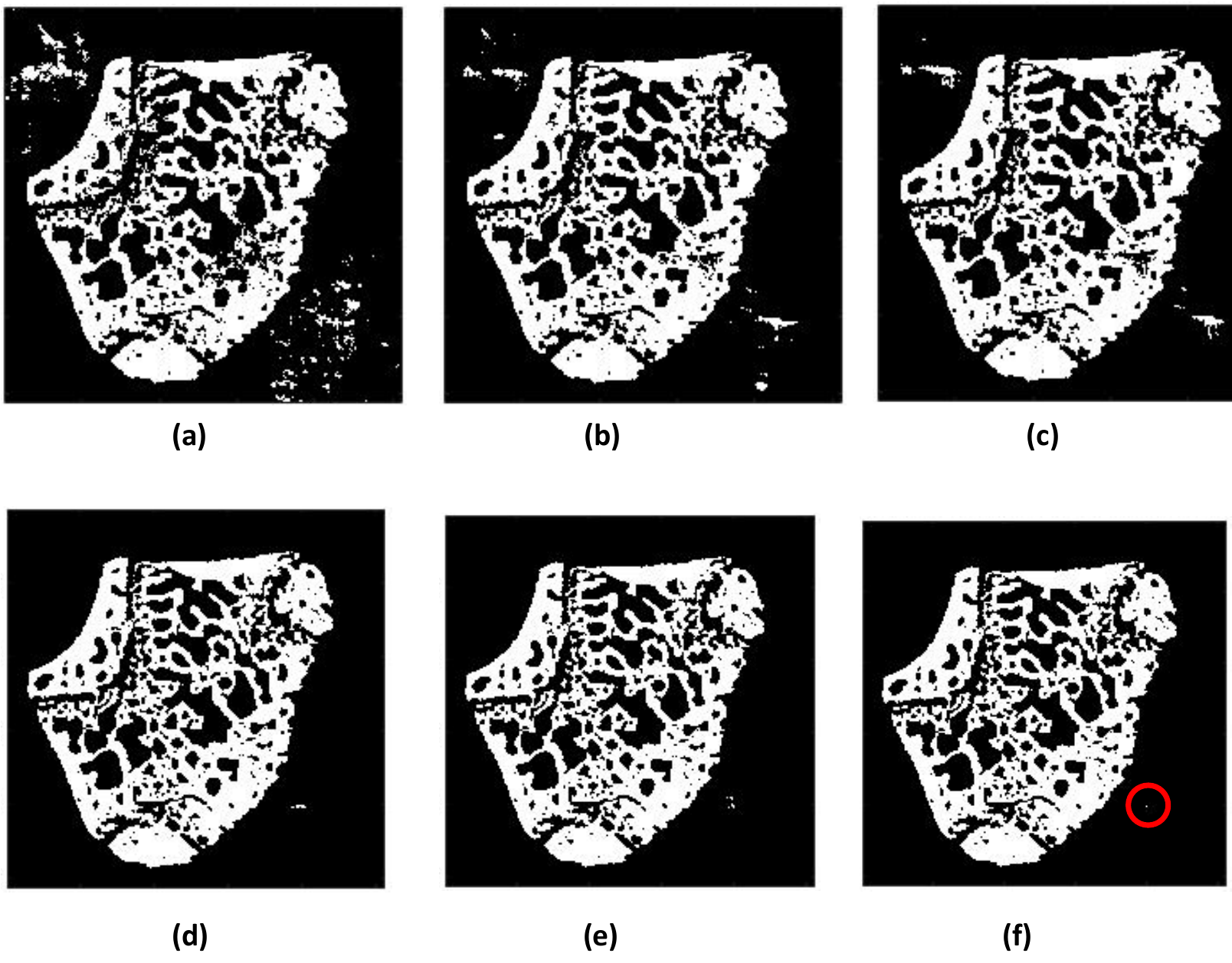}
\caption{Reconstruction of Phantom 4 by BRA for different numbers of iterations: (a) 10 iterations. (b) 50 iterations. (c) 100 iterations. (d) 250 iterations. (e) 400 iterations. (f) 500 iterations. The circle contains the very small region (12 pixels, see Table \ref{tab:wrongpixels}) where wrongly reconstructed pixels appear. Exact reconstruction is obtained within 550 iterations.}
\label{fig:phantom4}
\end{figure}

We have run BRA with different numbers of iterations and computed the corresponding percentages of reconstructions. Results are reported in Table \ref{tab:phantoms}, where, for each one of the four binary phantoms, the performances of pure CGLS and of BRA are compared. Figures \ref{fig:phantom1}-\ref{fig:phantom4} show different reconstruction outputs for different numbers of iterations.

\begin{table}[htbp]
\begin{tabular}{|l||c|c|c|c|c|c|c|c|c|c|c|c|}
\hline
\:\:\:\:\:\:\:\:\textbf{}&\multicolumn{3}{c|}{\textbf{Phantom 1}}&\multicolumn{3}{c|}{\textbf{Phantom 2}}&\multicolumn{3}{c|}{\textbf{Phantom 3}}&\multicolumn{3}{c|}{\textbf{Phantom 4}}\\\hline
&\multicolumn{3}{c|}{\textbf{$\%$ reconstruction}}&\multicolumn{3}{c|}{\textbf{$\%$ reconstruction}}&\multicolumn{3}{c|}{\textbf{$\%$ reconstruction}}&\multicolumn{3}{c|}{\textbf{$\%$ reconstruction}}\\\hline
\textbf{$\mathbf{\sharp}$ \textbf{iterations}}&\multicolumn{3}{c|}{\textbf{CGLS\hfill BRA}}&\multicolumn{3}{c|}{\textbf{CGLS\hfill BRA}}&\multicolumn{3}{c|}{\textbf{CGLS\hfill BRA}}&\multicolumn{3}{c|}{\textbf{CGLS\hfill BRA}}\\\hline\hline
$10$                                 &\multicolumn{3}{c|}{$83.38$\hfill$89.50$}&\multicolumn{3}{c|}{$85.66$\hfill $93.11$}&\multicolumn{3}{c|}{$88.71$\hfill $98.00$}&\multicolumn{3}{c|}{$86.95$\hfill $96.96$}\\\hline
$20$                                &\multicolumn{3}{c|}{$84.50$\hfill $92.90$}&\multicolumn{3}{c|}{$86.78$\hfill $94.50$}&\multicolumn{3}{c|}{$90.64$\hfill $99.10$}&\multicolumn{3}{c|}{$88.35$\hfill $98.20$}\\\hline
$30$                               &\multicolumn{3}{c|}{$85.80$\hfill $93.10$}&\multicolumn{3}{c|}{$87.72$\hfill $94.90$}&\multicolumn{3}{c|}{$92.18$\hfill $99.20$}&\multicolumn{3}{c|}{$89.20$\hfill $98.74$}\\\hline
$40$                               &\multicolumn{3}{c|}{$86.22$\hfill $93.20$}&\multicolumn{3}{c|}{$88.37$\hfill $95.20$}&\multicolumn{3}{c|}{$92.67$\hfill $99.30$}&\multicolumn{3}{c|}{$89.67$\hfill $98.86$}\\\hline
$50$                                 &\multicolumn{3}{c|}{$86.59$\hfill $93.30$}&\multicolumn{3}{c|}{$88.67$\hfill $95.40$}&\multicolumn{3}{c|}{$92.96$\hfill $99.50$}&\multicolumn{3}{c|}{$90.02$\hfill $99.03$}\\\hline
$100$                               &\multicolumn{3}{c|}{$88.20$\hfill $95.98$}&\multicolumn{3}{c|}{$90.02$\hfill $97.98$}&\multicolumn{3}{c|}{$93.90$\hfill $99.66$}&\multicolumn{3}{c|}{$91.18$\hfill $99.20$}\\\hline
$150$                                &\multicolumn{3}{c|}{$90.48$\hfill $97.95$}&\multicolumn{3}{c|}{$91.82$\hfill $99.20$}&\multicolumn{3}{c|}{$94.82$\hfill $99.75$}&\multicolumn{3}{c|}{$92.50$\hfill $99.50$}\\\hline
$180$                               &\multicolumn{3}{c|}{$91.90$\hfill $98.40$}&\multicolumn{3}{c|}{$92.82$\hfill $99.40$}&\multicolumn{3}{c|}{$95.18$\hfill $99.80$}&\multicolumn{3}{c|}{$93.36$\hfill $99.80$}\\\hline
$200$                             &\multicolumn{3}{c|}{$93.12$\hfill $99.10$}&\multicolumn{3}{c|}{$93.69$\hfill $99.60$}&\multicolumn{3}{c|}{$95.48$\hfill $99.84$}&\multicolumn{3}{c|}{$93.94$\hfill $99.96$}\\\hline
$250$                           &\multicolumn{3}{c|}{$95.61$\hfill $99.99$}&\multicolumn{3}{c|}{$95.88$\hfill $99.88$}&\multicolumn{3}{c|}{$96.50$\hfill $99.92$}&\multicolumn{3}{c|}{$95.00$\hfill $99.96$}\\\hline
$350$                                 &\multicolumn{3}{c|}{$96.53$\hfill $100$ }&\multicolumn{3}{c|}{$97.15$\hfill $99.97$}&\multicolumn{3}{c|}{$97.43$\hfill $99.99$}&\multicolumn{3}{c|}{$96.19$\hfill $99.95$}\\\hline
$400$                                 &\multicolumn{3}{c|}{$96.95$\hfill $100$ }&\multicolumn{3}{c|}{$97.51$\hfill $99.99$}&\multicolumn{3}{c|}{$97.60$\hfill $99.99$}&\multicolumn{3}{c|}{$96.48$\hfill $99.96$}\\\hline
$450$                                &\multicolumn{3}{c|}{$97.39$\hfill $100$  }&\multicolumn{3}{c|}{$97.82$\hfill $99.99$}&\multicolumn{3}{c|}{$97.72$\hfill $99.99$}&\multicolumn{3}{c|}{$96.69$\hfill $99.98$}\\\hline
$500$                                 &\multicolumn{3}{c|}{$97.80$\hfill $100$  }&\multicolumn{3}{c|}{$98.06$\hfill$100$ }&\multicolumn{3}{c|}{$97.81$\hfill $99.99$}&\multicolumn{3}{c|}{$96.89$\hfill $99.99$}\\\hline
$550$                                 &\multicolumn{3}{c|}{$97.97$\hfill $100$  }&\multicolumn{3}{c|}{$98.20$\hfill $100$  }&\multicolumn{3}{c|}{$97.87$\hfill $99.99$}&\multicolumn{3}{c|}{$97.10$\hfill$100$ }\\\hline
$600$                                 &\multicolumn{3}{c|}{$98.09$\hfill $100$  }&\multicolumn{3}{c|}{$98.31$\hfill $100$  }&\multicolumn{3}{c|}{$97.92$\hfill $99.99$}&\multicolumn{3}{c|}{$97.20$\hfill $100$  }\\\hline
$650$                                 &\multicolumn{3}{c|}{$98.18$\hfill $100$  }&\multicolumn{3}{c|}{$98.38$\hfill $100$  }&\multicolumn{3}{c|}{$97.96$\hfill$100$ }&\multicolumn{3}{c|}{$97.28$\hfill $100$  }\\\hline
\end{tabular}
\caption{Comparison between CGLS and BRA. Percentages of exact reconstruction of Phantoms 1-4 are shown for different numbers of iterations, until perfect reconstruction by BRA is obtained.}
\label{tab:phantoms}
\end{table}

As a further detail on the performance of BRA, in Table \ref{tab:wrongpixels} we report the number of wrongly reconstructed pixels for a few values of the number of iterations in the range $250-600$.

\begin{table}[htbp]
\begin{center}
\begin{tabular}{|l||c|c|c|c|c|c|c|c|c|c|c|c|}
\hline
\:\:\:\:\:\:\:\:\textbf{}&\multicolumn{3}{c|}{\textbf{Phantom 1}}&\multicolumn{3}{c|}{\textbf{Phantom 2}}&\multicolumn{3}{c|}{\textbf{Phantom 3}}&\multicolumn{3}{c|}{\textbf{Phantom 4}}\\\hline
$\mathbf{\sharp}$ \textbf{iterations}&\multicolumn{3}{c|}{\textbf{$\mathbf{\sharp}$ wrong}}&\multicolumn{3}{c|}{\textbf{$\mathbf{\sharp}$  wrong}}&\multicolumn{3}{c|}{\textbf{$\mathbf{\sharp}$ wrong}}&\multicolumn{3}{c|}{\textbf{$\mathbf{\sharp}$ wrong}}\\\hline\hline
$250$                                 &\multicolumn{3}{c|}{$21$}&\multicolumn{3}{c|}{$302$}&\multicolumn{3}{c|}{$210$}&\multicolumn{3}{c|}{$91$}\\\hline
$280$                                 &\multicolumn{3}{c|}{$15$}&\multicolumn{3}{c|}{$86$}&\multicolumn{3}{c|}{$135$}&\multicolumn{3}{c|}{$117$}\\\hline
$300$                                 &\multicolumn{3}{c|}{$9$}&\multicolumn{3}{c|}{$80$}&\multicolumn{3}{c|}{$88$}&\multicolumn{3}{c|}{$118$}\\\hline
$320$                                 &\multicolumn{3}{c|}{$5$}&\multicolumn{3}{c|}{$74$}&\multicolumn{3}{c|}{$71$}&\multicolumn{3}{c|}{$123$}\\\hline
$350$                                 &\multicolumn{3}{c|}{$0$}&\multicolumn{3}{c|}{$66$}&\multicolumn{3}{c|}{$36$}&\multicolumn{3}{c|}{$131$}\\\hline
$400$                                 &\multicolumn{3}{c|}{$0$}&\multicolumn{3}{c|}{$25$}&\multicolumn{3}{c|}{$12$}&\multicolumn{3}{c|}{$98$}\\\hline
$450$                                 &\multicolumn{3}{c|}{$0$}&\multicolumn{3}{c|}{$10$}&\multicolumn{3}{c|}{$12$}&\multicolumn{3}{c|}{$46$}\\\hline
$500$                                 &\multicolumn{3}{c|}{$0$}&\multicolumn{3}{c|}{$0$}&\multicolumn{3}{c|}{$12$}&\multicolumn{3}{c|}{$12$}\\\hline
$550$                                 &\multicolumn{3}{c|}{$0$}&\multicolumn{3}{c|}{$0$}&\multicolumn{3}{c|}{$12$}&\multicolumn{3}{c|}{$0$}\\\hline
$600$                                 &\multicolumn{3}{c|}{$0$}&\multicolumn{3}{c|}{$0$}&\multicolumn{3}{c|}{$4$}&\multicolumn{3}{c|}{$0$}\\\hline
\end{tabular}
\caption{Number of wrongly reconstructed pixels when iterations increase.}
\label{tab:wrongpixels}
\end{center}
\end{table}

\begin{figure}[htbp]
\centering
\includegraphics[scale=0.6, viewport=56 420 519 772, clip=true]{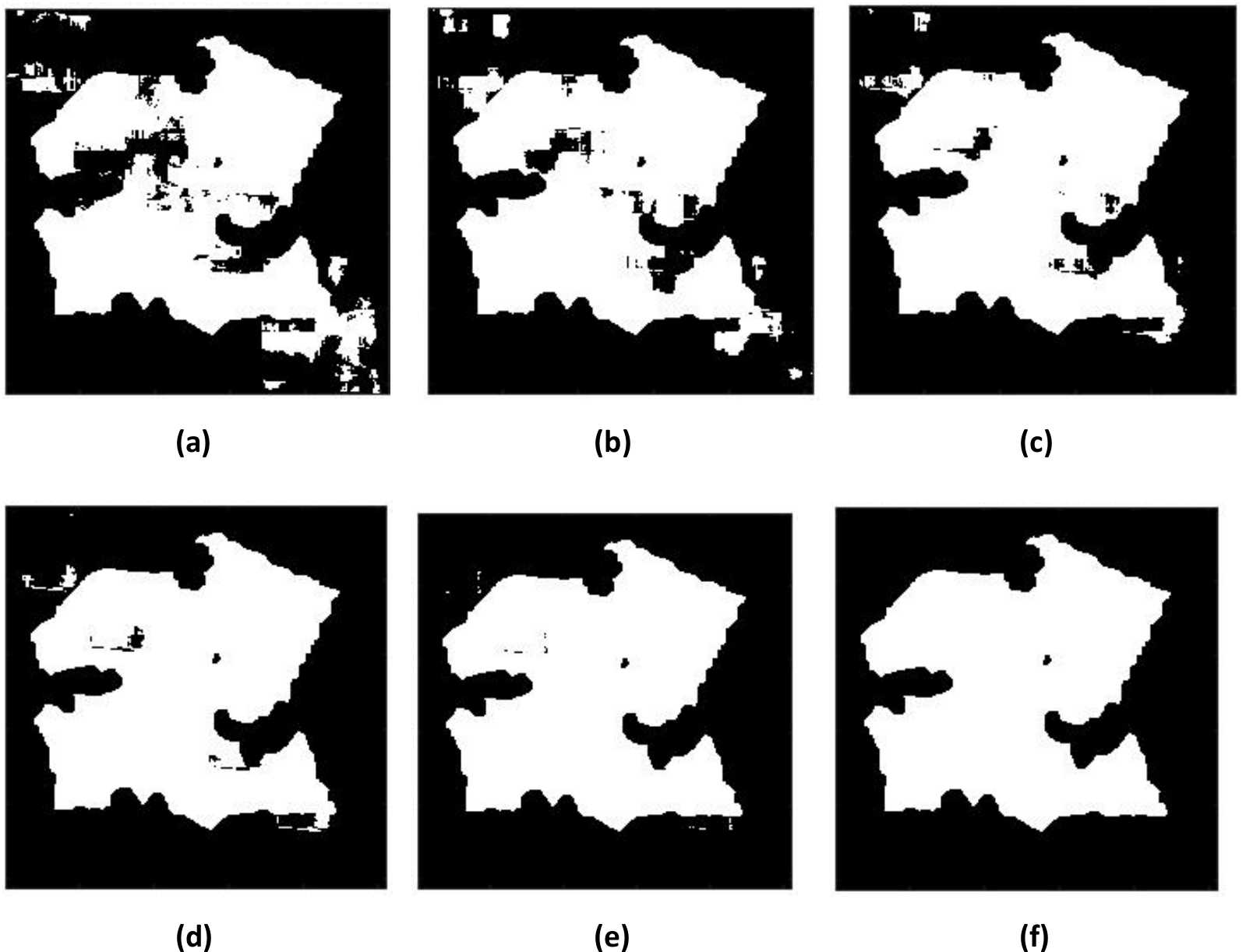}
\caption{Reconstruction of Phantom 2 by BRA for different numbers of iterations: (a) 10 iterations. (b) 50 iterations. (c) 100 iterations. (d) 150 iterations. (e) 250 iterations. (f) 400 iterations. Exact reconstruction is obtained within 500 iterations.}
\end{figure}

\begin{figure}[htbp]
\centering
\includegraphics[scale=0.6, viewport=56 420 519 772, clip=true]{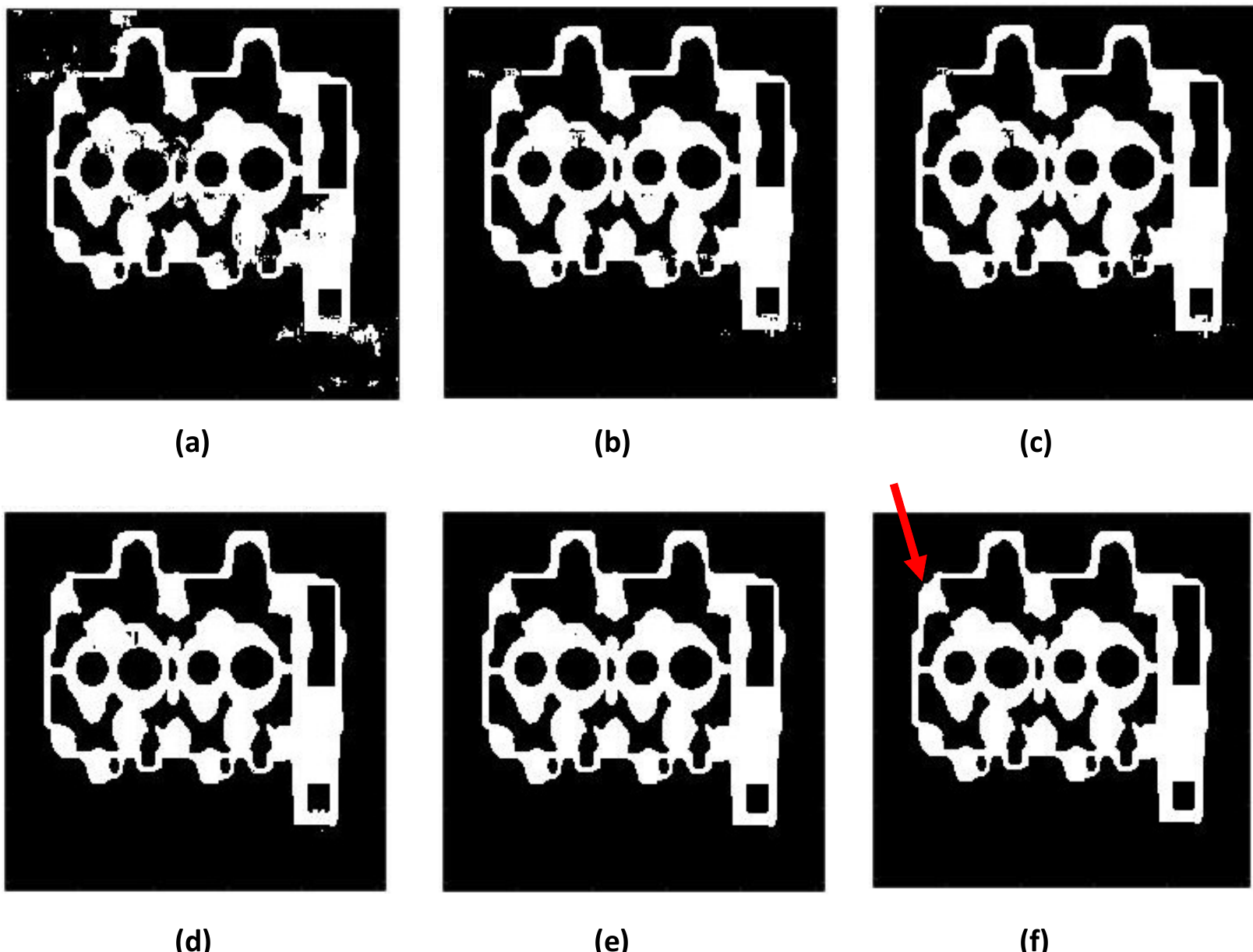}
\caption{Reconstruction of Phantom 3 by BRA for different numbers of iterations: (a) 10 iterations. (b) 50 iterations. (c) 100 iterations. (d) 200 iterations. (e) 400 iterations. (f) 600 iterations. The arrow points to the very small region (4 pixels, see Table \ref{tab:wrongpixels}) where wrongly reconstructed pixels appear. Exact reconstruction is obtained within 650 iterations.}
\end{figure}

\subsubsection{Discussion of the results}
Even though a large percentage of correctly reconstructed pixels is obtained within $100$ iterations, the complete reconstruction depends on the structure and complexity of the chosen phantom. As described in \cite{BFHT}, Phantom 1 represents a very simple object, with an almost smooth boundary, Phantom 2 shows an object with a very fragmented boundary and a small hole inside, Phantom 3 represents a cross-section of a cylinder head in a combustion engine and contains many holes, while Phantom 4 has been obtained from a micro-CT image of a rat bone.

Moreover, the performances of pure CGLS and of BRA have been compared. Even though the pure CGLS is able to reconstruct a considerable percentage of image, it always (namely, for each number of iterations) underperforms BRA, and never reaches exact reconstruction within the range of BRA, and beyond. When the image presents a large number of holes, for instance in case of Phantom 3 and Phantom 4, the number of iterations required by BRA to get exact reconstruction is $650$ and $550$, respectively. After the same number of iterations the pure CGLS exactly reconstructs a percentage of the image which is comparable with the results provided by BRA after just 10 iterations.

In the easiest case, namely Phantom 1, the complete reconstruction is obtained within $350$ iterations, while these are not enough for the other phantoms. However, note that more than $99\%$ of pixels of all phantoms are correctly reconstructed within just $200$ iterations. It seems that the last $1\%$ of pixels to be reconstructed requires the greatest effort in term of iterations, independently of the shape of the phantom. This is related to the total number of iterations that are required to complete the reconstruction process, which needs several other items in the case of Phantom 2, Phantom 3 and Phantom 4. Let us briefly comment on the reconstructions.

For Phantom 2 the great fragmentation of its boundary determines a considerable increased number of iterations with respect to Phantom 1, even if the last pixels to be reconstructed lie in the interior part. Note however that the small hole is immediately detected, namely, within the first 10 iterations.

It seems reasonable to relate the even larger number of iterations required for the complete reconstructions of Phantom 3 and Phantom 4 to their large number of holes. It is also interesting to observe that, in the case of Phantom 4, when the number of iterations ranges between $250$ and $350$ then a local increasing of wrongly reconstructed pixels appears. As commented in Section \ref{subsec:BRA_reconstructions}, this is caused by CGLS that, in this range of iterations, returns values close to $\frac{1}{2}$ for some pixels, which are alternatively rounded to either $0$ or $1$ until a suitable number of iterations is considered in order to stabilize the result. As an example, this occurs in pixel $(405,397)$, where it results
$$\begin{array}{l}
\mathbf{x}_{250}^{\ast}(405,397)=0.4789,\\
\mathbf{x}_{280}^{\ast}(405,397)=0.5021,\\
\mathbf{x}_{300}^{\ast}(405,397)=0.5045,\\
\mathbf{x}_{320}^{\ast}(405,397)=0.5077,\\
\mathbf{x}_{350}^{\ast}(405,397)=0.5061,\\
\mathbf{x}_{400}^{\ast}(405,397)=0.4890.
\end{array}$$
Pixel $(405,397)$ belongs to the region $\mathcal{A}\setminus H$, therefore BRA returns the binary rounding of $\mathbf{x}_{\kappa}^{\ast}(405,397)$ without any weight updating. Consequently, when $\kappa$ increases from $250$ to $280$, the value of $\overline{\mathbf{x}}_{\kappa}(405,397)$ changes from $0$ to $1$, then it remains $1$ for further increasing number of iterations until, for $\kappa\geq 400$, it returns definitively equal to $0$, which is precisely the value of Phantom 4 in the considered pixel.

Note that the number of iterations required to get exact reconstruction is $350$ for Phantom 1, $500$ for Phantom 2, $650$ for Phantom 3, $550$ for Phantom 4. The average is $512.5$, which supports the estimate $O(\sqrt{MN})$ ($M=N=512$ in these cases), as we have pointed out in Section \ref{subsec:BRA_reconstructions} when dealing with the complexity of BRA.

\section{Conclusion and comments}\label{sec:conclusion}
In this paper we have addressed the tomographic problem of finding an algorithm that provides exact noise-free reconstruction of a binary image in the grid model, in case a special set $S$ of four valid direction is employed. Starting from the uniqueness result of Theorem \ref{teo:uniqueness}, we have proved Theorem \ref{teo:rounding} and Corollary \ref{cor:roundingreconstruction}, which lead to Algorithm \ref{alg:BRA} (BRA). It works under a prescribed number $\kappa$ of iterations, and, when $\kappa$ is sufficiently large, BRA allows exact reconstruction of the unique binary solution in the grid model. The idea follows the same approach as in \cite{BFHT}, with the extra condition provided by Theorem \ref{teo:uniqueness}, which guarantees since the beginning that, if a set $S\in\mathcal{S}(\mathcal{A})$ is employed, then the solution in the lattice grid $\mathcal{A}$ exists and is unique. This allows the exact determination of the pixels belonging to the space of ghosts, and the consequent computation of their values in the image to be reconstructed by means of a binary rounding process on the entries of the real-valued solution of minimal Euclidean norm.

We have also explicitly implemented BRA and numerically tested its performance on the same binary phantoms considered in \cite{BFHT}. We have presented the results in two different tables, detailing how exact reconstructions are obtained in the four different cases.

Of course BRA, as presented here, is  mainly intended as an explicit implementation of the theoretical results provided in \cite{BDP1} (and related papers), so that it cannot be considered as an immediate counterpart to more sophisticated reconstruction algorithms (see for instance \cite{batsij} and the related bibliography). However, we think that the easy structure of BRA could be profitably matched with some usually employed strategies, in order to improve the speed and the quality of the reconstruction process.

In view of a reinforcement of the proposed approach, applications of BRA to real tomographic data would allow to actually investigate its pros and cons.  To this, a first step should be devoted to the adaptation of the grid model to a real tomographic acquisition system. In particular, it could be worth to test the robustness of BRA by replacing the lattice lines with strips of suitable width and working in the Dirac model (see Figure \ref{fig:stripmodel}). For a set $S$ of directions we could define its \emph{intrinsic width} as follows:

\begin{equation*}
w(S)=\min_{(a_r,b_r)\in S}\frac{1}{\sqrt{a_r^2+b_r^2}},
\end{equation*}
where the pixels have size $1\times 1$. The intrinsic width represents the minimal distance between two consecutive lattice lines having direction in $S$. In case we consider a width $w>w(S)$ then the projections collected in the grid model, and obtained from lines belonging to a same strip, are grouped together, so that the number of projections is lowered. The original linear system $A\mathbf{x}=\mathbf{p}_S$ ($\mathbf{x}\in\R^n$, $\mathbf{p}_S\in\R^m$) becomes of the form $A'\mathbf{y}=\mathbf{p}_S'$, where $\mathbf{p}_S'$ has size $s<m$. The new projection matrix $A'$ is obtained from $A$ by summing together different rows, corresponding to equations related to lines that fall in a same strip. Since only parallel lines are grouped, the unknowns appearing in the resulting equations are all distinct, namely, the matrix $A'$ is still binary. This suggests to interpret $A'\mathbf{y}=\mathbf{p}_S'$ as a linear system still associated to a grid model, in the same original lattice grid $\mathcal{A}$, but with a lower resolution. Consequently, for increasing $w$, the weakly bad configuration $F_S$ is enlarged accordingly, so that the structure of the $S$-ghost \eqref{eq:base_ghost} changes, and higher multiplicities may appear. This implies that the interval \eqref{eq:bounds} does not necessarily hold for all the involved parameters $\alpha^{\ast}_u$, and the central reconstruction $\mathbf{x}^{\ast}$ progressively moves away from the binary solution $\overline{\mathbf{x}}$. As a possible extension of the results presented in this paper, it would be worth investigating how the quality of reconstruction changes as $w$ increases.

Moreover, it would be desirable to modify BRA in order to include also the cases when noisy projections are considered. As a further extension, we wish to explore the same problems for gray-scale images. In this case, Theorem \ref{teo:uniqueness} is not valid anymore, so first of all a generalization of such a theoretical result to integer-valued images is needed.\smallskip

\textbf{Acknowledgments} The authors wish to thank the anonymous reviewers for their useful comments and valuable suggestions. The research of the second author has been partially supported by Fondazione Fratelli Confalonieri (http://www.fondazionefratelliconfalonieri.it/).

\section*{References}

\bibliographystyle{plain}
\bibliography{BTRWFP_171128}

\end{document}